\documentclass[lettersize,journal]{IEEEtran}

\usepackage{array}
\usepackage[caption=false,font=normalsize,labelfont=sf,textfont=sf]{subfig}
\usepackage{stfloats}
\usepackage{url}
\usepackage{verbatim}
\usepackage{cite}
\usepackage{amsmath,amssymb,amsfonts}
\usepackage{bbm}
\usepackage{graphicx}
\usepackage{textcomp}

\newcommand{\jfpackage}[1] {\usepackage{#1}}

\newcommand{\indef}[1] {\input{./#1}}
\jfpackage{jf-all}
\usepackage[table]{xcolor}

\definecolor{ForestGreen}{RGB}{34,139,34}
\usepackage{upgreek} 
\usepackage{hyperref}
\usepackage{cite}
\usepackage{booktabs}
\usepackage{enumerate}

\usepackage{algorithm} 
\usepackage{algpseudocode}

\newcommand{\codo}[1] {}
\long\def\red#1{\bgroup\color{red}#1\egroup}
\long\def\blue#1{\bgroup\color{black}#1\egroup}
\long\def\black#1{\bgroup\color{black}#1\egroup}
\long\def\magenta#1{\bgroup\color{magenta}#1\egroup}
\long\def\green#1{\bgroup\color{ForestGreen}#1\egroup}

\usepackage{graphicx} 
\usepackage{./support-caption}

\usepackage{todonotes}
\newcommand{\mycomment}[1]{}

\newcommand{\wiki}[1]	{\href{#1}{[wiki]}}

\newcommand{\tofref}[1]	{\todoa{Fig. (XXX)}}

\newcommand{\todof}[1]	{\todoa{Fig. (XXX)}}

\renewcommand{\defequ}	{\xmath{\triangleq}}

\newcommand{\veta}	{\blmath{\eta}}

\newcommand{\vrho}	{\blmath{\rho}}

\newcommand{\vh}	{\blmath{h}}
\newcommand{\vl}	{\blmath{l}}

\newcommand{\vs}	{\jfunop{\bm{s}}}

\newcommand{\vN}	{\blmath{N}}

\renewcommand{\reals}	{\jfunl{\mathbb{R}}}
\newcommand{\complex}	{\jfunl{\mathbb{C}}}
\newcommand{\nats}	{\jfunl{\mathbb{N}}}
\newcommand{\ints}	{\jfunl{\mathbb{Z}}}

\newcommand{\A}	{\blmath{A}}

\newcommand{\C}	{\blmath{C}}

\newcommand{\G}	{\blmath{G}}
\renewcommand{\H}	{\blmath{H}}
\newcommand{\I}	{\blmath{I}}

\renewcommand{\P}	{\blmath{P}}

\newcommand{\bS}	{\blmath{S}}

\newcommand{\V}	{\blmath{V}}
\newcommand{\W}	{\blmath{W}}

\newcommand{\Y}	{\blmath{Y}}

\newcommand{\norminf}[1]	{\xmath{\norm{#1}_{\infty}}}

\newcommand{\reg}	{\jfunl{\upbeta}}

\newcommand{\jcal}[1]	{\jfunl{\mathcal{#1}}}

\newcommand{\x}	{\blmath{\xsym}}

\newcommand{\nLsym}	{\ooalign{$\mathsf{L}$\cr\kern+0.0em\textsf{-}}}

\newcommand{\Rq}	{\pR_{\mathrm{Q}}}

\newcommand{\vd}	{\blmath{d}}

\renewcommand{\L} {\blmath{L}}
\renewcommand{\k} {\xmath{\vec{k}}}
\renewcommand{\vl} {\xmath{\vec{l}}}
\renewcommand{\x} {\xmath{\vec{x}}}

\newcommand{\Lpsf} {\xmath{L_{\mathrm{PSF}}}}

\renewcommand{\S}{\bS}

\newcommand{\mwrep} {\black{parsimonious }}
\newcommand{\missword} {applicability }
\newcommand{\Missword} {Applicability }

\newcommand{\kt} {\xmath{(\k, t)}}
\newcommand{\Gx} {\xmath{\G(\x)}}
\newcommand{\Jx} {\xmath{J(\x)}}
\newcommand{\Tbx} {\xmath{T_b(\x)}}
\newcommand{\Tblim} {\xmath{T_b}}
\newcommand{\Bx} {\xmath{\jcal B(\x)}}
\newcommand{\xf} {\xmath{(\x, f)}}
\newcommand{\trhokt} {\xmath{\Tilde{\rho}(\k, t)}}
 
\newcommand{\crhoxf} {\xmath{\check{\rho}(\x, f)}}
\newcommand{\rhoxt} {\xmath{\rho(\x, t)}}

\newcommand{\thkp} {\xmath{\Tilde{h}(\k, t)}}
\renewcommand{\rq} {\xmath{\black{\tau}}}
\renewcommand{\Rq} {\xmath{\black{P}}}
\newcommand{\Radq} {\xmath{\black{R}}}

\newcommand{\hq} {\xmath{h_{\rq}}}
\newcommand{\thq} {\xmath{\Tilde{h}_{\rq}}}
\newcommand{\pq} {\xmath{p}}
\newcommand{\thp} {\xmath{\Tilde{h}_{\pq}}}
\newcommand{\hpq} {\xmath{h_{\pq}}}
\newcommand{\thrkt} {\xmath{\thq(\k, t)}}
\newcommand{\thpkt} {\xmath{\thp(\k, t)}}
\newcommand{\hrxt} {\xmath{\hq(\x, t)}}

\newcommand{\rhoxf} {\xmath{\check{\rho}(\x, f)}}
\newcommand{\rankC} {\xmath{r_{\C}}}

\newcommand{\bxf} {\xmath{\check{b}(\x, f)}}
\newcommand{\ind} {\xmath{\mathbbm{1}}}
\newcommand{\bPhi}	{\blmath{\Phi}}

\newcommand{\tgvkt} {\xmath{\Tilde{g}_{t'}(\k, t)}}
\newcommand{\gvxt} {\xmath{g_{t'}(\x, t)}}
\newcommand{\cgvxf} {\xmath{\check{g}_{t'}(\x, f)}}
\newcommand{\LpS} {\xmath{\text{L $+$ S}}}
\newcommand{\stmLK}{\xmath{\text{STM \& P-LORAKS}}}
\newcommand{\stmTK}{\xmath{\text{STM \& Tikhonov}}}
\newcommand{\psfTK}{\xmath{\text{PSF \& Tikhonov}}}
\newcommand{\psfTKIV}{\xmath{\text{PSF \& Tikhonov ($L_{\text{PSF}} = 4$)}}}
\newcommand{\psfTKVI}{\xmath{\text{PSF \& Tikhonov ($L_{\text{PSF}} = 6$)}}}

\newcommand{\Tset} {\xmath{\{1, \ldots, T\}}}
\newcommand{\Tsum} {\xmath{\sum_{t=1}^T}}

\newcommand{\TTbset} {\xmath{\{-\floor{\Tblim/2} + 1,\ldots, T+\floor{\Tblim/2} \}}}

\begin{document}

\markboth{Journal of \LaTeX\ Class Files,~Vol.~XX, No.~X, Month~XXXX}%
{Shell \MakeLowercase{\textit{et al.}}: A Sample Article Using IEEEtran.cls for IEEE Journals}

\title{Spatiotemporal Maps for Dynamic MRI Reconstruction}

\author{Rodrigo A. Lobos, \IEEEmembership{Member, IEEE}, 
Xiaokai Wang,  
Rex T. L. Fung,  
Yongli He, 
David Frey,
Dinank Gupta, 
Zhongming Liu, \IEEEmembership{Senior Member, IEEE},
Jeffrey A. Fessler, \IEEEmembership{Fellow, IEEE},
and Douglas C. Noll, \IEEEmembership{Fellow, IEEE}
%
\thanks{
This work was supported by 
the National Institutes of Health through grants 
R21EB034344, R01EB035618
and 
R01AT011665. 
\textit{(Corresponding author: Rodrigo A. Lobos).}
}
\thanks{Rodrigo A. Lobos, 
Xiaokai Wang, 
Rex T. L. Fung, 
David Frey,
Dinank Gupta,
Zhongming Liu,
and Douglas C. Noll
are with the Department of Biomedical Engineering,
University of Michigan, Ann Arbor, MI 48109 USA
(e-mail: 
rlobos@umich.edu,
xiaokaiw@umich.edu,
rexfung@umich.edu,
djfrey@umich.edu,
dinankg@umich.edu,
zmliu@umich.edu,
dnoll@umich.edu).
}
\thanks{
Yongli He
is with the Department of Applied Physics, 
University of Michigan, Ann Arbor, MI 48109 USA
(e-mail: yonglihe@umich.edu).
}
\thanks{
Jeffrey A. Fessler
is with the Department of Electrical Engineering and Computer Science, 
University of Michigan, Ann Arbor, MI 48109 USA
(e-mail: fessler@umich.edu).
}
\thanks{
This paper has supplementary downloadable material available 
at http://ieeexplore.ieee.org., provided by the author. 
The material includes additional tables, figures, and videos. 
This material is 15 MB in size.
}

}

\maketitle

\begin{abstract}

The partially separable functions (PSF) model 
is commonly adopted in dynamic MRI reconstruction,
as is the underlying signal model in many reconstruction methods
including the ones
relying on low-rank assumptions.
Even though the PSF model offers a \mwrep representation
of the dynamic MRI signal in several applications, 
its representation capabilities tend to decrease
in scenarios where voxels present different temporal/spectral characteristics
at different spatial locations. 
In this work we account for this limitation
by proposing a new model, 
called spatiotemporal maps (STMs),
that leverages autoregressive properties of (k, t)-space.
The STM model decomposes the spatiotemporal MRI signal
into a sum of components, 
each one consisting of a product between
 a spatial function and a
temporal function that depends on the spatial location.
The proposed model can be interpreted as an extension of the PSF model
whose temporal functions 
are independent of the spatial location.
We show that spatiotemporal maps can be efficiently computed
from autocalibration data
by using advanced signal processing
and randomized linear algebra techniques, 
enabling STMs to be used
as part of many reconstruction frameworks 
for accelerated dynamic MRI. 
As proof-of-concept illustrations,
we show that STMs can be used 
to reconstruct both
2D single-channel animal gastrointestinal MRI data
and 3D multichannel human functional MRI data.
\end{abstract}


\section{Introduction}
\label{sec:introduction}

\IEEEPARstart{D}{ata}-acquisition below the Nyquist rate
is desired in dynamic MRI 
to achieve high spatial and temporal resolution.
In the multichannel case, 
considering a $Q$-channel receiver array, 
sub-Nyquist (accelerated) data are often modeled 
as samples of the continuous spatial Fourier transform as:
\be
    d_q(\k, t) = \int c_q(\x)\rho(\x,t)e^{-i2\pi\k \cdot \x} \der \x
    + \eta(\k, t),
\ee{e,dyn_mri_sc}
where 
$d_q(\k, t)$ denotes a complex-valued sample in the \kt-space 
from the $q$th coil receiver, $q\in\{1,\ldots Q\}$;
$c_q(\x)$ 
is a complex-valued sensitivity map for the $q$th coil receiver; 
each
$\k \in \reals^D$
and 
$t \in \reals$
denotes a k-space location 
and time point,
respectively,  
at which the complex-valued underlying dynamic signal $\rho(\x, t)$
is sampled in the \kt-space; 
$D \in \{2, 3\}$ is the dimension of the image; 
$\x \in \reals^D$ denotes a spatial location; 
and $\eta(\k, t)$ denotes the 
white complex Gaussian noise observed in the sample. 
In accelerated scans, 
data are collected during a finite number of time frames $T$, 
and $M$ k-space samples are acquired in each time frame
out of an ideal total of $N$ samples 
needed for a reconstruction according to the Nyquist theorem. %
After acquiring the accelerated data,
the goal is to reconstruct
the image sequence $\rho(\x,t)$ 
from the $QMT$ \kt-space measurements.

Many reconstruction methods
have been proposed to solve this inverse problem
\cite{ 
jones1993k,
liang1997dynamic,
madore1999unaliasing,
liang:07:siw,
tsao2003k,
jung2009k,
haldar2010spatiotemporal,
lingala2011accelerated,
zhao:12:irf,
trzasko:11:lvg,
lingala2013blind,
otazo:15:lrp,
feng2016xd,
poddar2015dynamic,
ravishankar:17:lra,
lin:19:edp,
biswas2019dynamic,
qin2018convolutional,
ke:21:llr,
cruz2023low,
lobos2025smooth,
feng2025spatiotemporal,
yu2025bilevel
},
including methods
that use partially separable function (PSF) models
\cite{liang:07:siw}.
PSF models approximate the spatiotemporal signal $\rho(\x,t)$ 
as a sum of products, 
each one consisting of
a temporal function
$\varphi_l(t)$
and a spatial function
$\rho_l(\x)$%
:
\be
\rho(\x, t) \approx \sum_{l = 1}^{\Lpsf} \varphi_l(t)\rho_l(\x).
\ee{e,psf}
The PSF model provides a parsimonious decomposition for $\rho(\x,t)$
in cases where temporal/spectral characteristics
are shared
across different spatial locations.
Notably, in these cases it can be theoretically shown
that a Casorati matrix constructed from the data
has low-rank characteristics 
\cite{liang:07:siw,
haldar2010spatiotemporal}
that have been leveraged  
by many dynamic MRI reconstruction methods
based on low-rank models
\cite{haldar2010spatiotemporal,
lingala2011accelerated,
trzasko:11:lvg,
zhao:12:irf,
otazo:15:lrp,
lin:19:edp,
cruz2023low
}.
On the other hand, 
the number of components in the PSF decomposition
must be large in cases where 
a high degree of variability is observed
for the temporal/spectral characteristics
across different spatial locations,
decreasing the effectiveness
of low-rank methods.
One way to overcome this limitation 
is to divide the spatial locations into multiple patches
and 
assume that
the Casorati matrix for each patch
has low-rank characteristics,
which is the principle behind locally low-rank (LLR) methods
\cite{
trzasko:11:lvg,
manjon2013diffusion,
saucedo2017improved,
tamir:17:tss,
cordero2019complex,
guo:20:hro,
vizioli:21:ltt, 
zhao2022joint,
comby:23:dof, 
meyer:23:ect, 
chen:23:isl,
cruz2023low,
lobos2025smooth
}. 
However, implementing LLR methods 
involves further challenges:
patches should overlap
to avoid blocky artifacts,
and the patch size should be chosen
assuming that the underlying PSF decomposition
for each patch is parsimonious
\cite{
saucedo2017improved,
lobos2025smooth},
which might not hold always. 
This work examines
if it is possible to provide
a parsimonious decomposition in a smaller scale than patches.
More specifically, if different parsimonious decompositions
can be provided for different spatial locations, 
such that temporal/spectral characteristics 
at different spatial locations could be captured separately. 

We propose a signal model of the form
\be
\rho(\x, t) \approx \sum_{l = 1}^{L(\x)} s_l(\x,t) \, \rho_l(\x),
\ee{e,stm}
where we call the temporal functions
$s_l(\x,t), ~ l = 1,\ldots, L(\x)$ 
\textit{spatiotemporal maps} (STMs)
and the number of components
can depend on the spatial location.
We call the decomposition in \eref{e,stm}
the STM model. 
The STM model has the potential to provide 
a more parsimonious representation than PSF models,
as dynamic behaviors that substantially vary
between spatial locations
are modeled separately.
This is equivalent to 
STM needing fewer components than PSF,
which can be leveraged in reconstruction settings. 

The first question that we theoretically explore in this work 
regards the scenarios in which the STM model, 
as presented in \eref{e,stm}, 
is applicable.
We show that the \missword of spatiotemporal maps
is based on establishing the existence of 
shift-invariant linear-predictability (SILP) relationships
in \kt-space
\cite{haldar:20:lpi}.
Moreover, we show that
a sufficient condition for the existence
of these SILP relationships, 
is that the time series for each 
spatial location
should have a spectral support in the \xf-space
mostly limited to a finite set of frequency bands,
\ie, such that $\rho(\x,t)$ could be modeled
as a multiband signal \cite{mishali2009blind}, 
where the set of frequency bands is not restricted
to be the same across spatial locations.

The second question that we address is 
how to efficiently calculate spatiotemporal maps
from the acquired accelerated data. 
Recent work in 
subspace-based sensitivity map estimation in multichannel MRI
\cite{lobos2023new},
has shown that the existence of
SILP relationships across the k-space of different channels 
is closely related to the estimation of sensitivity maps. 
By considering the different time frames
as virtual channels in a parallel imaging setting,
and showing that SILP relationships exist 
across the k-space of different time frames, 
we show how to calculate spatiotemporal maps
using analogous principles 
to the ones used in 
subspace-based sensitivity map estimation.
Notably, this allows us to use 
recently developed computational methods
for the estimation of sensitivity maps
to efficiently compute spatiotemporal maps
from autocalibration data.
Specifically, we draw inspiration from PISCO 
\cite{lobos2023new,
lobos2023extended,
lobos2023software},
a set of computational methods
that has enabled fast sensitivity map estimation.

In many dynamic MRI applications
the number of time frames is
much larger than the number of coils
in typical receiver arrays,
so some of the original PISCO techniques 
are slower than
in sensitivity map estimation.
This issue is exacerbated in 3D data.
One of the most demanding tasks in PISCO
is calculating a singular value decomposition (SVD)
of a big Hankel/Toeplitz convolutional matrix 
whose dimensions increase 
with the number of channels,
which corresponds to the number of time frames 
in the dynamic MRI setting.
Here we extend the PISCO method by adding
a randomized linear algebra technique 
that calculates a proxy for this SVD
using sketching \cite{gilbert2012sketched}.
This technique significantly reduces computation time,
enabling STM calculation
for long time series
while negligibly affecting representation quality.
One PISCO technique
uses power iterations
to estimate a single nullspace vector 
for a family of Hermitian matrices.
Here we extend this technique 
using orthogonal iteration \cite[p.~454]{golub2013} 
instead of a power iteration, 
because calculating spatiotemporal maps
requires a nullspace basis
for each of the aforementioned Hermitian matrices,
rather than just
one nullspace vector.

Finally, we show that the STM model in \eref{e,stm}
can be easily incorporated into a reconstruction framework
for accelerated dynamic MRI.
Inspired by reconstruction methods
based on the PSF model
\cite{zhao:12:irf}, 
our strategy consists of 
computing the spatial functions 
in the STM model,
\ie, $\{\rho_l(\x)\}_{l=1}^{L(\x)}$, 
from undersampled \kt-space,
assuming that the number of components $L(\x)$
per voxel
is much smaller than the number of frames.
Using spatiotemporal maps 
that we precompute from autocalibration data,
the final reconstructed time series
uses the model \eref{e,stm}.
This approach can be complemented
by adding regularizers for the spatial functions
in cases where the data is heavily undersampled.
We explore the combination of the STM model 
with several regularizers,
and we evaluate this approach 
in two challenging  dynamic MRI datasets:
2D single-channel animal gastrointestinal MRI data
and 
3D multichannel human functional MRI (fMRI) data.
A preliminary version of this work was presented as a short abstract \cite{lobos2025stm_ismrm}.

The formulation of the STM model in \eref{e,stm}
resembles the one used in the blind compressed sensing (BCS) method 
proposed in \cite{lingala2013blind}.
However, the principles guiding the learning
of the voxel-dependent temporal functions are distinct.
In BCS, a dictionary of temporal functions 
is learned from the undersampled \kt-space data 
using an energy constraint based on the Frobenius norm.
Then, \rhoxt is modeled as a linear combination of the dictionary temporal functions,
with voxel-dependent coefficients chosen via sparsity constraints.
In the STM model,
the temporal functions (\ie, the spatiotemporal maps)
are learned without constructing a dictionary.
Instead, the STMs are learned independently for each voxel 
by applying a subspace-based method to autocalibration data,
that rely on SILP relationships in \kt-space.

This paper is organized as follows.
Section \ref{sec:theory} presents the theory
behind the \missword of spatiotemporal maps 
for dynamic MRI reconstruction. 
Section \ref{sec:est_stm} reviews
the computational methods 
used to compute spatiotemporal maps, 
and introduces novel techniques 
that extend PISCO. 
Section \ref{sec:rec} provides 
a reconstruction framework 
for accelerated dynamic MRI 
using spatiotemporal maps. 
Section \ref{sec:exp} describes the experiments that
evaluate the proposed STM model
and reconstruction framework,
and Sec.~\ref{sec:res} provides the corresponding results.
Lastly, Sec.~\ref{sec:dis_con} provides 
discussion and final remarks.

\section{Theory}
\label{sec:theory}

This section establishes a theoretical framework  
for the applicability of
spatiotemporal maps.
We observe 
that calculating spatiotemporal maps
is closely related to 
estimating sensitivity maps in parallel imaging. 
In fact, by setting $L(\x) = 1$ in \eref{e,stm}
and considering time frames as virtual channels,
the STM model recovers 
the commonly used parallel imaging model%
\footnote{
By setting $L(\x)>1$ 
it is possible to recover the case
when more than one set of sensitivity maps is needed,
for example, when aliasing exists 
due to a small field-of-view 
\cite{uecker2014,
lobos2023new
}.
}.
In this case,
spatiotemporal maps become sensitivity maps.
We extend the theoretical framework
proposed in \cite{lobos2023new}
for subspace-based sensitivity map estimation
to calculating spatiotemporal maps. 
This theoretical framework starts by assuming
that SILP relationships exist across channels
in a parallel imaging scenario
\cite{haldar:20:lpi,
lobos2023new,
lobos2023extended
}. 
Therefore,
here we show that SILP relationships
can also exist across time frames
(for both single coil and multichannel MRI)
in the dynamic MRI setting.
For this purpose, we first
formally introduce the concept of SILP relationships in \kt-space
and study sufficient conditions for their existence.

\subsection{Shift-invariant Linear Predictability Relationships in
\texorpdfstring{\kt-space}{kt-space}} \label{sec:th_silp}

We assume that 
the spatial support of \rhoxt is finite for all time,
\ie,
that the set of spatial locations
$\Om \defequ \{\x \in \reals^D : \exists \, t\in \reals, \, |\rhoxt|>0 \}$
is completely contained within a hypercube \Gam,
corresponding to the field-of-view (FOV).
We normalize the coordinates so that each FOV width is unity.
In other words, the object being imaged
remains within the FOV over time.
Analogously, 
after normalizing the temporal frequency range,
we assume that \rhoxt is
approximately bandlimited
to the set
$\jcal F = [-\frac{1}{2}, \frac{1}{2}]$
for all $\x \in \Gam$.
In other words,
$\rhoxf \approx 0, ~\forall f\notin \jcal F$,
where \rhoxf denotes
the 1D temporal Fourier transform%
\footnote{
We use
$\check{\gamma}(\x, f)$ 
and 
$\Tilde{\gamma}(\k, t)$
to denote 
the temporal Fourier transform
and 
the spatial Fourier transform
of the function $\gamma(\x, t)$,
respectively.
The variables \k and $t$
can be continuous or discrete,
as should be clear from the context.
}
of \rhoxt:
\be
\rhoxf \defequ \int \rhoxt \, e^{-i2\pi ft} \der t.
\ee{e,temp_ft}
These assumptions imply
the following 
Fourier series representations:
\begin{align}
\rhoxf &\approx \ind_{\jcal F}(f) \sum_{t \in \ints} \rho(\x, t)
\, e^{-i2\pi f t},
\label{e,fs_t}
\\
\rho(\x, t) &= \ind_{\Gam}(\x) \sum_{\k \in \ints^D} \trhokt
\, e^{i2\pi \k \cdot \x},
\label{e,fs_k}
\end{align}
where 
\rhoxt in \eref{e,fs_t} 
denotes temporal samples 
of the dynamic MRI signal 
according to the Nyquist rate
based on $\jcal F$%
;
\trhokt
denotes samples of 
the spatial Fourier transform of \rhoxt,
located on a rectilinear Nyquist grid
with k-space sample spacing given by
a sampling period $\Del k = 1$ 
(in each dimension),
which follows from our FOV assumption;
and
$\ind_{\jcal U}(u)$ is the indicator function of the set $\jcal U$
which is equal to $1$ if $u\in\jcal U$ or equal to $0$ otherwise.

In real applications,
we only have access
to samples in the \kt-space
for a finite number of time frames $T$.
We index the time samples using the set
\Tset.

An SILP relationship across the \kt-space
of the $T$ available time frames 
is equivalent to
the existence of a nonzero multiframe finite impulse response (FIR) filter
\thkp
that satisfies:
\be
\Tsum\sum_{\vl\in \Lambda}
\Tilde{h}(\vl, t) \, \Tilde{\rho}(\k - \vl, t) \approx 0
, \quad \forall \k \in \ints^D,
\ee{e,silp_kt}
where 
$\Lambda\subset \ints^D$ 
is the finite support set of each
$\thkp,  \, t \in \Tset,$
in k-space. 
If \eref{e,silp_kt} holds,
then
there is an annihilation relationship
among the k-space samples of different time frames
that is shift-invariant \cite{haldar:20:lpi}.
The following subsection
provides sufficient conditions for the existence
of filters satisfying \eref{e,silp_kt}.
These conditions are based on the assumptions that 
\rhoxt has a limited support in the \xf-space
located in a small set of disjoint bands,
and that its spectral characteristics 
smoothly vary across the FOV.
These assumptions 
provide sufficient conditions for the existence 
of SILP relationships as in \eref{e,silp_kt};
however, we are not providing necessary conditions. 
There might be other cases where SILP relationships could be established.

\subsection{Sufficient Conditions for the Existence of SILP Relationships
in the \kt-space} \label{sec:th_scond}

So far we have assumed that 
the frequency support of \rhoxf
is restricted to the same band
$\jcal F = [-\frac{1}{2}, \frac{1}{2}]$
for every spatial location in the FOV.
However,
it is possible to provide more refined modeling assumptions
when each spatial location is considered separately,
as this enables more parsimonious models
where \rhoxt exhibits distinct behavior
across spatial locations in the FOV. 
Hereafter
we assume that
for every $\x \in \Gam$, 
\rhoxf is further restricted to be
(approximately)
supported on no more than
$\Jx \in \nats$
disjoint intervals (bands) in $\jcal F$,
which is equivalent to assuming that 
\rhoxt is a multiband signal
\cite{mishali2009blind}. 
We denote the location-dependent union of these bands by 
$\Bx \subset \jcal F$.
We further assume that the set of bands
smoothly varies across spatial locations in the FOV,
akin to locally low-rank models that assume
\rhoxt exhibits similar behaviors
for nearby spatial locations.

Under these support conditions,
one can define a nonzero function \bxf 
that is spatially smooth, 
bandlimited to $\jcal F$,
and that for each $\x \in \Gam$
its frequency support is mostly defined on 
$\jcal F \setminus \Bx$,
where ``$\setminus$'' denotes the set difference.
Such a function satisfies:
\be
\bxf \, \rhoxf \approx 0,
\quad \forall f \in \jcal F, \, \forall \x\in\Gam.
\ee{e,an_fil_f}
Under our on-going assumptions
there are many functions 
that satisfy \eref{e,an_fil_f}.
This approximation could be arbitrarily close to zero
if we allow \bxf to have arbitrary rapid variations in frequency
in cases where the number of bands \Jx is relatively large;
however, such variations would complicate
computing \bxf in practice. 
Thus,
we restrict our analysis to annihilator functions
(\ie, functions that satisfy \eref{e,an_fil_f}) 
that are smooth in $f$,
\ie, 
whose temporal support
is concentrated in a small time interval.
Specifically, 
we require that the 
1D inverse discrete time Fourier transform (IDTFT) of \bxf
given by
\be
b(\x, t) \defequ \frac{1}{2}\int_{-\frac{1}{2}}^{\frac{1}{2}} \bxf
\, e^{i2\pi f t} \der f
, \quad t\in\ints,
\ee{e,g_idtft}
is approximately zero when
$|t - \floor{T/2} - 1| > \floor{\Tbx/2}$, 
where $\Tbx\in \nats$ can depend on the spatial location
and $\floor{\cdot}$ denotes the flooring operation.
For simplicity,
hereafter we assume that
$b(\x, t) \approx 0$ when $|t - \floor{T/2}-1| > \floor{\Tblim/2}$,
where
$\Tblim \defequ \max_{\x \in \Gam} \Tbx$.
It follows from \eref{e,an_fil_f}
and the DTFT convolutional theorem 
that
\be
\sum_{t\in \ints}  \rhoxt \, b(\x, \rq - t) \approx 0, \quad \forall \rq \in \ints,
\ \forall \x\in\Gam.
\ee{e,conv_dt}
If $T$ is large enough such that 
\be
\rhoxf \approx \ind_{\jcal F}(f)\Tsum \rhoxt \, e^{-i2\pi f t},
\ee{e,rhoxf_ap}
then \eref{e,conv_dt} implies that
\be
\Tsum \hq(\x,t) \, \rhoxt \approx 0, \, \forall \x\in\Gam,
\ee{e,conv_dt_ap}
for $\rq \in \TTbset$, 
where we define
$\hq(\x, t) \defequ b(\x, \rq - t)$
and for simplicity
we assume
$T$ and \Tblim are both odd.
Next, we apply the Fourier transform convolution theorem
on the spatial domain to obtain that
\be
\Tsum \sum_{\vl\in\Lambda}
\thq(\vl, t) \, \Tilde{\rho}(\k-\vl, t) \approx 0,
\quad \forall \k \in \ints^D,
\ee{e,conv_sp}
$\rq \in \TTbset$, 
where 
\thrkt,
the spatial discrete Fourier transform of 
\hrxt,
has its support $\Lambda$ restricted 
by our assumption on the spatial smoothness of \bxf.
Finally, each function 
\thrkt
corresponds to a multiframe FIR filter
satisfying \eref{e,silp_kt}.

The previous analysis shows 
that many FIR filters satisfying \eref{e,silp_kt} 
can be found for one function \bxf;
however, infinitely many other functions
can be defined with the same characteristics of \bxf.
This property suggests that multiple SILP relationships 
should exist in \kt-space across time frames
when \rhoxt follows the multiband model 
at each spatial location. 
Our theoretical analysis 
is inspired by similar observations
in previous work 
\cite{haldar:14:lrm,
haldar:20:lpi,
lobos2023extended}.

Our previous analysis 
assumes that the set of bands where \rhoxf
is nonzero varies smoothly across the FOV.
However, this condition may not hold in real applications.
If \rhoxf is not spatially smooth, 
then \bxf should be defined accordingly
to ensure that the approximation in \eref{e,an_fil_f} remains valid.
Allowing \bxf to vary rapidly across the FOV
can improve the approximation in \eref{e,an_fil_f},
and enable multiframe FIR filters to achieve a better approximation in \eref{e,conv_sp}.
However, these filters would require a larger support $\Lambda$, 
making their computation more challenging in practice (\cf, Sec. III.A).
The next subsection shows that 
choosing the filter support size 
involves a trade-off between computational complexity
and representation accuracy of the STM model.

The following subsection first shows
how SILP relationships
connect with the \missword of spatiotemporal maps.

\subsection{SILP Relationships in \kt-space and the \Missword of Spatiotemporal Maps}
\label{sec:SILP_stm}

Recent work in multichannel MRI
has shown that the existence of SILP relationships 
across channels in a parallel imaging setting
is closely related to the estimation of
sensitivity maps using subspaces 
\cite{lobos2023new}. 
Given that 
SILP relationships can also exist across time frames in \kt-space,
this section shows
how to extend the theoretical framework in \cite{lobos2023new}
to STM calculation.

Section \ref{sec:th_scond} showed that
many multiframe FIR filters exist that satisfy \eref{e,silp_kt} 
by assuming a multiband behavior for \rhoxt at each spatial location.
Hereafter we assume 
that $\Rq$ of these filters are available,
denoted by
$\thpkt, ~ \pq \in \{1,\ldots, \Rq\}$.
Therefore, we have $\Rq$ equations 
following the same structure as \eref{e,silp_kt}.
Using analogous derivations to the ones in \cite{lobos2023new},
which involve the Fourier transform convolutional theorem,
we equivalently express these $\Rq$
equations in the spatial domain 
using matrix-vector multiplications.
By using the spatial representation 
\be
\hpq(\x, t) \defequ \sum_{\k \in \Lambda}\thpkt \, e^{i2\pi \k \cdot \x},
\quad t\in \Tset,
\ee{e,fir_xt}
it follows that
\be
\H(\x) \, \vrho(\x) \approx \bm 0,  \quad \forall \x\in \Gam,
\ee{e,silp_xt_mv}
where
$\H(\x) \in \complex^{\Rq \times T}$ 
has its $(\pq,t)$th entry defined as 
$[\H(\x)]_{\pq t} \defequ \hpq(\x, t)$,
and 
$\vrho(\x)\in\complex^T$
has elements
$[\vrho(\x)]_t \defequ \rho(\x, t)$.
In words,
\eref{e,silp_xt_mv}
says that $\vrho(\x)$ 
is an approximate nullspace vector of $\H(\x)$.
Because the number of filters $\Rq$ 
tends to be much larger than
the number of frames $T$, 
it is more convenient 
to use the nullspace relationship
\be
\G(\x) \, \vrho(\x) \approx \bm 0,  \quad \forall \x\in \Gam,
\ee{e,silp_xt_mv_G}
where
$\G(\x) \defequ \H^H(\x) \H(\x) \in  \complex^{T\times T}$,
whose $(t',t)$th entry is given by
\be
[\G(\x)]_{t' t} = \sum_{\pq=1}^\Rq \hpq^*(\x, t') \, \hpq(\x, t),
\ee{e,Gxt't}
where $(\cdot)^*$ denotes the complex conjugate operation.
In the k-space domain,
\eref{e,silp_xt_mv_G} is equivalent to 
reducing the initial \Rq SILP equations 
of the form
\eref{e,silp_kt}
to $T$ SILP equations of the form 
\be
\Tsum\sum_{\vl\in \Theta}
\Tilde{g}_{t'}(\vl, t) \, \Tilde{\rho}(\k - \vl, t) \approx 0
, \quad \forall \k \in \ints^D,
\ee{e,silp_kt_P}
$t'\in\{1,\ldots, T\}$, where 
\be
\tgvkt \defequ \sum_{\pq = 1}^\Rq \Tilde{h}^*_{\pq}(-\k,t') \conv \thp(\k,t);
\ee{e,g_filt}
here $\conv$ denotes k-space domain convolution;
and $\Theta$ is the finite support 
of the multiframe filters $\{\tgvkt\}_{t' =1}^T$.

The annihilation property
\eref{e,silp_xt_mv_G} suggests that
$\vrho(\x)$ could 
be approximated by a scaled version of
the eigenvector of $\G(\x)$
with the smallest eigenvalue,
or by some linear combination of a set of eigenvectors
whose corresponding eigenvalues are all zero or nearly zero.
We call such eigenvectors the ``nullspace''
of $\G(\x)$,
though in practice the smallest eigenvalues
might not be exactly zero
so ``very small space''
might be a more apt name.
In other words,
\eref{e,silp_xt_mv_G}
suggests that we can represent $\vrho(\x)$
(approximately)
using a basis for the nullspace of $\G(\x)$.
Therefore,
as detailed in Sec.~\ref{sec:est_stm},
we compute spatiotemporal maps by first computing
an orthonormal basis for the nullspace of $\G(\x)$.

For each \x,
let
$\vs_l(\x) \in \complex^T, ~ l\in\{1, \ldots, L(\x)\},$
denote orthonormal vectors composing a basis 
for the nullspace of $\G(\x)$,
where $L(\x)$ is the nullspace dimension.
Then we propose the following signal model:
\be
\vrho(\x) \approx \sum_{l=1}^{L(\x)}\vs_l(\x)\rho_l(\x), 
\quad \forall \x\in \Gam,
\ee{e,rho_basis}
where
$\{\rho_l(\x)\}_{l=1}^{L(\x)} \subset \complex$
are scalar coefficients
that must be computed
by an image reconstruction algorithm.
We define
spatiotemporal maps (STMs)
as follows:
\be
s_l(\x, t) \defequ [\vs_l(\x)]_t, 
\quad \forall \x\in \Gam,
\ee{e,stm_def}
for  $l\in\{1, \ldots, L(\x)\}$
and $t \in \Tset$.

The signal decomposition provided in \eref{e,rho_basis}
has all its elements depending on the spatial location,
including the number of components $L(\x)$. 
We selected this parameter by studying  
the eigenvalues of the \Gx matrices (\cf, \fref{fig:eig_svd}). 
For each spatial location $L(\x)$ can be
the number of eigenvalues below a user-selected threshold. 
However,
for simplicity,
for the empirical results in Sec.~\ref{sec:exp}
we used the same number of components
for each spatial location.
Specifically,
we set the number of components for each spatial location to
$L \defequ \max_{\vec{x}\in\Gam} L(\vec{x}), \, \forall \x\in \Gam $,
and we relied on the scalar coefficients
$\rho_l(\x)$ to be zero in cases 
where the number of components was overestimated.
Having the number of components 
depend on the spatial location 
could help further reduce degrees of freedom
in future work.

The accuracy of the approximation in \eref{e,rho_basis}
depends primarily on how well the basis 
for the (approximate) nullspace of $\H(\x)$ can be computed.
Additionally, the validity of the nullspace relationship in \eref{e,silp_xt_mv} plays a significant role. 
This latter relationship, in turn, heavily depends 
on how closely the multiframe FIR filters satisfy
the SILP relationships in \eref{e,silp_kt}.
If these conditions are not met--meaning
the approximation in \eref{e,silp_kt} deviates significantly from zero--then the approximation in \eref{e,rho_basis} will also be less accurate,
ultimately decreasing the representation capabilities of the STM model. 
Therefore, accurately computing multiframe FIR filters
is an important step in calculating STMs.

Our derivation 
assumes that $P$ multiframe FIR filters exist that satisfy \eref{e,silp_kt}.
Our theoretical analysis in Sec.~\ref{sec:th_scond}
indicates that the accuracy of the approximation in \eref{e,silp_kt}
may depend on the size of the support $\Lambda$. 
Specifically, 
increasing the size of the support $\Lambda$ 
can lead to a better approximation in \eref{e,silp_kt},
which in turn improves the nullspace approximation in \eref{e,silp_xt_mv} 
throughout the FOV.
However, a larger support $\Lambda$ can also allow for 
more rapid variations in the spatial domain of the multiframe FIR filters. 
As a consequence, the matrix $\H(\x)$ may change more rapidly from voxel to voxel.
This leads to the nullspace of each matrix $\H(\x)$
varying more quickly across the FOV,
resulting in more rapid changes in their bases.
Consequently, the resulting STMs are less smooth.
Additionally, 
as we show in the following section,
computing multiframe FIR filters with a larger support
increases computational complexity of calculating STMs (\cf, Sec.\ref{sec:est_stm_1}).

\section{Computing Spatiotemporal Maps}
\label{sec:est_stm}

This section summarizes 
the practical steps for computing spatiotemporal maps efficiently
from autocalibration k-space data.
Each of these steps
relies on PISCO \cite{lobos2023new}, 
a set of computational methods 
originally proposed
for efficient subspace-based sensitivity map estimation in multichannel MRI. 
We show that PISCO can be adapted
for computing spatiotemporal maps
due to the theoretical connections
with sensitivity map estimation 
shown in previous sections.
For simplicity,
we first present
the steps for computing STMs
for the single-channel case, 
leaving the extension to the multichannel case
to the end of this section.

\subsection{Step 1: Calculating multiframe FIR filters} \label{sec:est_stm_1}

Assuming that SILP relationships exist in \kt-space,
the first step in computing spatiotemporal maps
corresponds to calculating 
$\Rq$ multiframe FIR filters satisfying \eref{e,silp_kt}.
Let $\Tilde{\vh}_{\pq}\in\complex^{|\Lambda|T}$
denote the vectorized version
of one of these multiframe filters.
One can rewrite \eref{e,silp_kt}
in matrix-vector form
to show
that each $\Tilde{\vh}_{\pq}$ is
an approximate nullspace vector of the matrix \cite{haldar2016p}
\be
\C \defequ \begin{bmatrix}
    \C_{1} & \C_{2} & \cdots & \C_{T}
\end{bmatrix} \in \complex^{I\times |\Lambda|T},
\ee{e,c_matrix}
where 
$\C_t \in \complex^{I \times |\Lambda|}$
is a matrix with a Hankel/Toeplitz convolutional structure
constructed from
\trhokt,
\ie,
from
k-space samples from the $t \, $th time frame,
with each of its $I$ rows corresponding
to a vectorized neighborhood of k-samples 
of size $|\Lambda|$ where \eref{e,silp_kt} holds
\cite{
haldar:14:lrm,
haldar2016p,
haldar:20:lpi,
lobos2023new}.
The type of  linear relationship described in \eref{e,silp_kt} 
is shift-invariant in k-space; 
consequently, any neighborhood with size $\Lambda$ 
available in k-space could be used to form a row of \C.
In practice,
we construct \C
from autocalibration data that is Nyquist-sampled 
and has sufficiently large enough dimensions
to allow the extraction of k-space neighborhoods of size $\Lambda$ from it.
We consider autocalibration data from the center of k-space 
as these samples commonly exhibit higher signal-to-noise ratio 
than samples from outer regions.
Then, we use an SVD to compute
an orthonormal basis for its (approximate) nullspace.
This basis,
denoted
$\vN \in \complex^{|\Lambda|T \times \Rq}$,
is expected to generate
the subspace of multiframe filters 
with support $\Lambda$ that satisfy \eref{e,silp_kt}.
Two options for the 
FIR filter support $\Lambda$
were studied
in \cite{lobos2023new}
when computing \C:
an ellipsoidal shape 
\ie,
$\Lambda = \{\k \in \ints^D ~ : ~ \normr{\k}_2 \leq \Radq \}$,
and
a rectangular shape 
\ie, 
$\Lambda = \{\k \in \ints^D ~ : ~ \normr{\k}_{\infty} \leq \Radq \}$,
where $\norm{\cdot}_2$ and $\norminf{\cdot}$
denote the $\ell_2$ and infinity norms, respectively.
The experiments in Sec.~\ref{sec:exp}
used FIR filters with an ellipsoidal shape
that saves computation
without sacrificing representation quality
compared to using a rectangular shape
\cite{lobos:22:ots,
lobos2023new}.
This choice is particularly relevant
when $D=3$.

As discussed in previous sections, 
using multiframe FIR filters with a larger support $\Lambda$
can improve the representation accuracy of the STM model; particularly,
when the data characteristics in the \xf-space
vary non-smoothly across the FOV.
This suggests that choosing a larger value for $R$ may be beneficial.
However, increasing $R$ also enlarges the dimensions of \C,
which raises memory usage
and computation time,
especially when using SVD 
to compute an orthonormal basis for its nullspace.
Therefore, the selection of $R$
poses a trade-off between representation capabilities of the STM model
and computational complexity. 
Furthermore, 
\C is computed using 
k-space neighborhoods of size $|\Lambda|$
extracted from autocalibration data.
If $R$ is set too large relative to the size of the available autocalibration data,
only a limited number of k-space neighborhoods could be extracted. 
This restriction reduces the number of rows in \C, 
which may compromise the accuracy with which its nullspace can be computed.

The dimensions of \C tend to be undesirably large,
which affects memory usage and computation time 
when computing a basis for its nullspace using SVD. 
In many applications $I > |\Lambda|T$,
so it is preferable to work with $\C^H\C$ instead of \C,
as they share the same nullspace.
PISCO provides FFT-based computational methods
to efficiently calculate an approximation of $\C^H\C$ 
without calculating \C first,
by leveraging the convolutional structure
of \C;
for details see
\cite[Sec. IV.A]{lobos2023new}.

\subsection{Step 2: Calculating \texorpdfstring{$\G(\x)$}{G}
for each spatial location}
\label{sec:est_stm_2}

After computing $\Rq$ multiframe FIR filters from \C,
the next step is to calculate $\G(\x)$ for each $\x\in \Gam$.
Evaluating \eref{e,fir_xt}
for each $\x\in \Gam$
and then computing
$\G(\x) = \H(\x)^H \H(\x)$
would be computationally expensive
when the FOV is big.
Fortunately, 
one of the computational methods provided in PISCO
can
directly calculate $\G(\x)$ using an FFT-based approach
without forming $\H(\x)$ first;
see 
\cite[Sec. IV.C]{lobos2023new}
for details.
In general terms, 
this method expresses
the entries of \Gx
using the k-space representation of the $\Rq$ FIR multiframe filters.
Specifically, this equivalent expression
uses the entries of the matrix
$\W \defequ \sum_{\pq = 1}^{\Rq} \Tilde{\vh}_{\pq} \Tilde{\vh}_{\pq}^H$.
Because step 1) computes the filters
$\{\Tilde{\vh}_{\pq}\}_{\pq=1}^{\Rq}$ 
as nullspace vectors of the \C matrix
and correspond to the columns of the matrix \vN,
it follows that $\W = \vN \vN^H$.
Therefore, we solely need $\vN$ from step 1)
to calculate $\G(\x)$ for each $\x\in \Gam$.

\subsection{Step 3: Construction of spatiotemporal maps}

The last step is to construct the spatiotemporal maps
using \eref{e,stm_def} by calculating
a basis for the nullspace of \Gx for each $\x \in \Gam$. 
A natural approach would be to use an SVD for each matrix $\G(\x)$,
which would be computationally expensive.
When the nullspace dimension of \Gx is equal to one
-- which is usually the case found in
subspace-based sensitivity map estimation --
PISCO provides a method based on power iteration
to compute the nullspace vectors
for all the spatial locations simultaneously
\cite[Sec.~IV.E]{lobos2023new}.
For dynamic imaging,
we expect to need $L(\x) > 1$,
\ie,
STMs require
nullspace bases with more than one vector.
Thus,
here we used
the orthogonal iteration \cite[p.~454]{golub2013}\cite{uecker2014}
instead of the power iteration. 
This approach computes nullspace bases
for all spatial locations simultaneously
for a general number of basis vectors
(\ie, the number of components $L$).

The aforementioned steps and computational techniques
enable efficient STM computation
in many cases. 
However, computational efficiency can be substantially improved 
by assuming that STMs are spatially smooth.
Under this assumption,
we can compute spatiotemporal maps
on a grid with coarser resolution
than the original
and then interpolate to the desired resolution.
PISCO also provides an FFT-based method for this procedure
that we directly use to compute STMs;
see \cite[Sec.~IV.D]{lobos2023new}
for details.
Nevertheless, 
the degree of interpolation should be properly selected,
as smoothness of STMs can be reduced 
when choosing a large value for $R$,
as previously discussed.

\subsection{Sketched SVD for the Projection onto the Nullspace
of \texorpdfstring{\C}{C}.} 
\label{sec:est_stm_sketch}

Computing STMs
is particularly challenging when $D=3$ 
and $T$ is large,
because finding \vN in step 1)
involves an SVD of 
$\C^H\C \in \complex^{|\Lambda|T \times |\Lambda|T}$, 
whose dimensions become inconveniently large in such cases.
When $D=3$, using multiframe FIR filters 
with an ellipsoidal shape support 
reduces $|\Lambda|$ considerably 
compared to using a rectangular shape support \cite{lobos:22:ots}. 
However, these savings are still insufficient
when $T$ is large,
\ie, for long time series (\eg, fMRI).
Given that \vN is used solely to calculate $\vN\vN^H$ in step 2),
here we propose a novel technique 
based on randomized linear algebra
to efficiently approximate $\vN\vN^H$
without calculating the SVD of $\C^H\C$.

Our approach uses a sketched SVD 
\cite{gilbert2012sketched}, 
a randomized linear algebra method,
to efficiently calculate approximations
for the singular values and singular vectors
of a large matrix.
Applied to our case, 
we propose to use a sketched SVD of the matrix $\C^H\C$.
Sketched SVD uses a random matrix 
that satisfies the Johnson-Lindenstrauss property 
\cite[p.~110]{vershynin2018high},
called the sketch matrix.
We denote this matrix by 
$\bPhi \in \complex^{s \times |\Lambda|T}$,
where $1 \leq s \ll |\Lambda|T$
denotes the sketch dimension.
Using the theoretical results from \cite{gilbert2012sketched},
it is possible to choose $s$
greater than the rank of \C,
such that the singular values and singular vectors
of the matrix $\Y \defequ \bPhi \C^H\C$
are similar to those of $\C^H\C$.
If $\rankC$ denotes the rank of \C,
it is commonly observed that $\rankC \ll |\Lambda|T$,
because the matrix \C has low-rank characteristics
\cite{haldar:14:lrm,
haldar2016p,
haldar:20:lpi}. 
By choosing the sketch dimension such that
$\rankC < s \ll |\Lambda|T$,
calculating an SVD of \Y
requires less computation
than calculating an SVD of $\C^H\C$.
If the sketch dimension is appropriately chosen \cite{gilbert2012sketched},
the right singular vectors of \Y 
approximate the ones of $\C^H\C$. 
However, this approximation 
would hold only for the \emph{first} $\rankC$
right singular vectors \cite{gilbert2012sketched}.
Because \vN is a \emph{nullspace} basis,
we need approximations for the \emph{last}
$|\Lambda|T - \rankC$
right singular vectors,
which are not provided
by the traditional sketched SVD.

Fortunately it is not
necessary to find \vN
to compute the projection matrix
$\vN\vN^H$.
Instead we construct an approximation for
$\vN\vN^H$
from the matrix whose columns are
the \emph{first} \rankC right singular vectors of \Y,
that we denote by 
$\Tilde\V\in \complex^{|\Lambda|T \times \rankC}$.
Using a sketched SVD to compute
$\Tilde\V \approx \V$,
where 
$\V \in \complex^{|\Lambda|T \times \rankC}$
is the matrix whose columns are
the first $\rankC$ right singular vectors of $\C^H\C$,
it follows from the fundamental theorem of linear algebra
\cite{strang:93:tft},
\cite[p.~181]{fessler2024linear}
that
\begin{align}
    \vN\vN^H &= \I_{|\Lambda|T} - \V\V^H
    \\
    & \approx \I_{|\Lambda|T} - \Tilde\V\Tilde\V^H, \label{eq:proj_ap}
\end{align}
where $\I_{|\Lambda|T}$ is the 
$|\Lambda|T \times |\Lambda|T$
identity matrix.

The authors in \cite{gilbert2012sketched} provide sufficient conditions
to select the sketch dimension $s$
such that an accurate singular vector approximation holds with high probability.
However, we have observed in our experiments 
that this selection rule can provide 
a sketch dimension that is too large 
to significantly reduce computation time.
In view of this, we provide a heuristic approach to select $s$.
We start by calculating a regular SVD of $\C^H\C$
considering a small number of time frames, 
such that the SVD computation is
less computationally demanding.
Then, we make a first estimation of \rankC
by selecting the point where the singular value curve
(sorted in decreasing order)
starts to flatten out. 
Using this estimation of \rankC,
we select the sketch dimension $s\in [2\rankC, 6\rankC]$,
and proceed to estimate the projector matrix using \eref{eq:proj_ap}. 
A subsequent refined estimation of \rankC can be made
using the singular values of \Y.
We found that this heuristic approach 
significantly reduces computation time
while negligibly affecting STM representation quality
(\cf, \fref{fig:time_NPR}).

\subsection{Spatiotemporal Map Computation for Multichannel Data}
The previous computation process
assumes that an autocalibration region 
sampled at the Nyquist rate (\ie, ACS data)
is acquired in each time frame.
In the single-channel case, 
we immediately proceed with the STM computation
using the ACS data to calculate $\C^H\C$
in the first step given in Sec. \ref{sec:est_stm_1}.
However, in the multichannel case, 
we apply a preliminary step 
to construct a virtual single-channel dataset 
from the ACS data.
In our experiments using multichannel data
we performed a SENSE-based coil combination in each time frame 
using sensitivity maps \cite{pruessmann:99:sse},
after zero-padding everything outside the ACS data. 
This creates a single-channel low-resolution dataset
that we used subsequently for spatiotemporal map computation.

\section{Reconstruction of undersampled dynamic MRI data
using spatiotemporal maps.} 
\label{sec:rec}

This section shows how to use \eref{e,stm}
to reconstruct accelerated dynamic MRI data. 
Assuming a Cartesian grid with $N$ points, 
we rewrite \eref{e,stm} in a vectorized form as
$
\vrho \approx \sum_{l=1}^L\S_l\vrho_l,
$
where $\vrho\in\complex^{NT}$ 
is  the vectorized version of $\rhoxt$ 
when considering all the samples and time frames;
$\S_l\in\complex^{NT\times N}$ 
is the matrix version of $s_l(\x, t)$,
which consists of a stack of $T$ diagonal matrices 
where each one contains the spatiotemporal map entries of one time frame; 
and $\vrho_l\in\complex^{N}$ 
is the vectorized version of $\rho_l(\x)$.
Then,
we rewrite the signal model in \eref{e,dyn_mri_sc} as:
\be
\vd \approx \A \left(\sum_{l=1}^L\S_l\vrho_l\right) + \veta,
\ee{e,dyn_mri_sc_mv}
where 
$\vd\in\complex^{QMT}$
is the vectorized version 
of the available samples in the \kt-space considering all the coils;
$\veta\in\complex^{QMT}$
is the vectorized version of the noise present in each sample;
and 
$\A\in\complex^{QMT\times NT}$
is a system matrix that includes
multiplication with sensitivity maps,
spatial Fourier transform of each time frame,
and an undersampling operation that is time-frame dependent.
The reconstruction problem
now becomes estimating
$
\vrho = \sum_{l=1}^L \S_l \vrho_l
$
from $\vd$,
given \A
and given the STMs $\{\S_l\}_{l=1}^L$
computed in Sec.~\ref{sec:est_stm}.
One could apply any of numerous reconstruction methods
from the literature to the measurement model
\eref{e,dyn_mri_sc_mv}.

For the experimental results in Sec.~\ref{sec:exp},
we simply focused on model-based image reconstruction approaches
where
one estimates $\vrho$ as follows:
\be
\hat{\vrho} = \sum_{l=1}^L\S_l \hat{\vrho}_l.
\ee{e,est_rec}
Here
\begin{align}
\{\hat{\vrho}_1,\ldots, \hat{\vrho}_L\} 
= 
\argmin{\{\vrho_l\}_{l=1}^L \subset \complex^{N}}
&\frac{1}{2}\norm{\A\left(\sum_{l=1}^L\S_l\vrho_l\right)- \vd}_2^2
\nonumber
\\
\nonumber \\
&+
\reg \, \jcal R(\vrho_1, \ldots, \vrho_L),
\label{e,rec_inv}
\end{align}
$\reg$
denotes a regularization term,
and $\jcal R$ corresponds to a regularizer
that imposes prior assumptions
on the spatial functions of the STM decomposition.
The STM formulation 
reduces the reconstruction of a dynamic time series
to computing a few ``static'' images, 
where the temporal dynamics are encoded in the spatiotemporal maps.
Therefore, we can use standard regularizers
developed for non-dynamic MRI such as 
Tikhonov, 
total variation + $\ell_1$, 
or regularizers based on structured low-rank models 
\cite{
haldar:14:lrm,
kim2017loraks
},
among numerous possible options.
Moreover, because this formulation includes 
the spatiotemporal maps in a data consistency term,
it could potentially be combined
with machine learning reconstruction methods
developed for non-dynamic MRI.

The following section 
explores the representation capabilities of the STM decomposition
and
tests the illustrative reconstruction methods
on realistic dynamic MRI data.

\section{Experiments}
\label{sec:exp}

We start describing the MRI datasets used in our experiments.
All animal procedures followed a protocol approved by the 
Unit for Laboratory Animal Medicine and the Institutional
Animal Care \& Use Committee at the University of Michigan.
For human subjects, 
approval for all ethical and experimental 
procedures and protocols 
was granted by the
University of Michigan Medical School Institutional Review Board.

\subsection{Data Description}
\begin{itemize}
    \item[(A.1)]{\textbf{Fully sampled 2D single-channel animal data}.}
    Gastrointestinal \textit{in vivo} MRI data of a rat 
    was acquired on a 7T single-coil small-animal scanner
    using a T1-weighted contrast and a gradient echo sequence.
    Gating was used to reduce respiratory motion.
    The slice orientation allowed visualizing
    movement of the rat's stomach 
    during digestion. 
    The data was acquired using a $128 \times 84$ Cartesian grid 
    where the dimensions corresponded to 
    the readout (RO) and phase encoding (PE) directions, respectively;
    $100$ time frames were acquired 
    with a temporal resolution of $\sim 1.6$ secs.
    Fully sampled data were acquired in each time frame.
    Further details of the acquisition can be found in \cite{wang2023diffeomorphic}.
    Supplementary video S1 visualizes the time frames.

    \item[(A.2)] {\textbf{Prospectively accelerated 2D single-channel animal data}.} 
    Dataset with the same acquisition characteristics as dataset (A.1).
    In this case the Cartesian grid had dimensions $128 \times  48$
    and the data was prospectively accelerated 
    using an undersampling mask 
    that in each time frame contained a
    $128 \times 12$ ACS region in the center of the k-space,
    and $12$ evenly spaced PE lines 
    outside the ACS region.
    These PE lines were shifted at different time frames 
    such that each k-space location was sampled
    at least once every $3$ time frames.
    Therefore, the undersampling mask produced  $\times 2$ 
    accelerated data. 
    For illustrative purposes our results using this dataset
    were zero-padded in the k-space domain
    to match the dimensions of dataset (A.1) (\ie, $128 \times 84$).
    Supplementary video S2 visualizes the time frames.

    \item[(B)]{\textbf{Fully sampled 3D multichannel human BOLD fMRI data}.} 
    BOLD fMRI data was acquired on a 3T MRI scanner
    using a 32-coil receiver array 
    and 
    a 3D gradient-echo EPI sequence. 
    A right-hand finger tapping task was performed
    with alternating $20$ secs blocks of tap/rest for 5 cycles.
    The data was acquired using a Cartesian grid 
    with dimensions $90 \times 90 \times 20$ 
    corresponding to 
    the RO direction,
    the PE direction in $k_y$,
    and the PE direction in $k_z$,
    respectively,
    with a $2.4 \, \text{mm}^3$ isotropic resolution;
    $140$ time frames were acquired 
    with a temporal resolution of $\sim 1.6$~secs.
    Fully sampled data were acquired in each time frame.
    The data were coil-compressed to $10$ virtual coils
    using a standard SVD approach \cite{buehrer:07:acf}.
\end{itemize}

All datasets presented their own challenges
for accelerated reconstruction.
On the one hand, 
only a single channel was available for datasets (A.1) and (A.2); 
therefore, no parallel imaging models could be used.
In addition, they exhibited a high level of noise
(\cf, supplementary videos S1 and S2),
and the temporal/spectral characteristics varied considerably
at different spatial locations.
On the other hand,
dataset (B) was 3D, which increased computation time 
for both spatiotemporal map computation and reconstruction.
In addition, the temporal variations of interest in fMRI
are inherently small in magnitude.
Finally, there were many time frames 
for both datasets,
so
the proposed 
sketched SVD approach in Sec.~\ref{sec:est_stm_sketch}
was crucial for practical
spatiotemporal map computation.

All datasets exhibited a dynamic behavior 
that was quasi-periodic for most spatial locations,
that also varied not too rapidly across the FOV.
Therefore, according to our theoretical analysis in Sec. \ref{sec:th_scond},
we expected to have a good approximation using the STM model.

\subsection{Undersampling k-space Masks
\label{sec:exp_mask}
for Retrospective Data Acceleration}

\begin{figure}[t] 
\centering 
\includegraphics[width=0.4\textwidth]{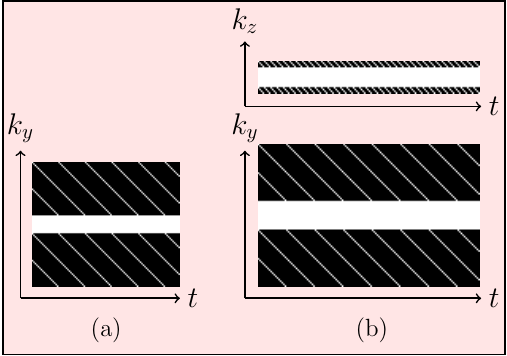}
\caption{
(a) k-space mask used to retrospectively undersample 2D+T dataset (A.1). 
White indicates the PE lines considered in each time frame. 
(b) Analogous k-space mask used to retrospectively undersample 3D+T dataset (B).
In this case two PE directions are considered.
}
\label{fig:masks}
\end{figure}

For our reconstruction experiments
we simulated accelerated acquisitions
by retrospectively undersampling the 
previously described datasets (A.1) and (B).
For dataset (A.1) we used an undersampling mask
that in each time frame contained 
a $128 \times 12$ ACS region in the center of the k-space,
and $4$ evenly spaced PE lines 
outside the ACS region.
These PE lines were shifted at different time frames 
such that each k-space location was sampled
at least once every $18$ time frames.
Therefore, the undersampling mask produced  $\times 5.25$ 
accelerated data.
Figure~\ref{fig:masks}.a illustrates
this undersampling mask.

For dataset (B) the undersampling mask was designed analogously.
It contained a $90 \times 18 \times 12$ ACS region 
in the center of the k-space of each time frame,
and $4$ PE lines in the $k_y$ direction outside the ACS region.
In addition, 
this mask included PE lines in the $k_z$ direction
outside the ACS region. 
Specifically, $2$ lines were considered.
Therefore, this mask produced 
$\times 5.84$
accelerated data.
Figure \ref{fig:masks}.b illustrates
this undersampling mask.
Though this sampling mask 
produced heavily undersampled data,
the acquisition time would not be considerably decreased
in a prospectively accelerated scenario, 
as the EPI sequence rapidly traverses the full $(k_x, k_y)$-space.
Our goal for investigating this undersampling mask
was to assess, as a proof-of-concept for fMRI, the STM model 
and the proposed reconstruction framework
in a heavily undersampled scenario.
Our goal in future work is to use the STM model
for other undersampled 3D k-space patterns for fMRI, 
for example, \cite{xiang2024model}.

\subsection{Implementation Details for Computing Spatiotemporal Maps}
For all experiments we computed spatiotemporal maps
using the steps in Sec. \ref{sec:est_stm}.
For datasets (A.1) and (A.2), 
in step 1) we used the ACS region 
of the undersampling mask
described in Sec. \ref{sec:exp_mask},
and we calculated $\C^H\C$ using
the FFT-based approach provided in PISCO
with an ellipsoidal neighborhood of radius $\Radq = 3$ 
for $\Lambda$ (\cf, Sec. \ref{sec:est_stm_1}).
Step 2) computed
a projection matrix for the nullspace of \C
using the sketched SVD approach
proposed in Sec. \ref{sec:est_stm_sketch}
with
a complex-valued Gaussian matrix for the sketch matrix $\bPhi$.
We selected the sketch dimension $s =  2r_{\C}$
using the heuristic approach described in Sec. \ref{sec:est_stm_sketch}.
Finally, step 3)
computed spatiotemporal maps
by calculating an approximate nullspace basis for each $\G(\x)$ matrix
using 
an SVD performed for each spatial location in the FOV.
If not specified otherwise,
we used $L = 4$ components.
To further reduce computation time,
we computed spatiotemporal maps
on a grid with coarser resolution
than the original 
and interpolated 
using the FFT-based PISCO approach.
The $12.5\%$ of each spatiotemporal map was interpolated in k-space
if not specified otherwise.
For dataset (B) the procedure was analogous.
The main differences with respect to datasets (A.1) and (A.2) were: the ACS region
corresponded to the one considered in the undersampling mask
for dataset (B) described in Sec. \ref{sec:exp_mask}; 
we selected $\Radq = 4$ 
if not specified otherwise;
and we used the orthogonal iteration to 
find the nullspace basis of each matrix $\G(\x)$.

\subsection{Representation Capabilities: Comparison with the PSF Model}

We compared the representation capabilities of the STM model
against the PSF model using dataset (A.1).
We computed spatiotemporal maps 
and the temporal functions of the PSF decomposition 
using the same ACS data,
and we calculated the normalized projection residual (NPR)
for both models while varying the number of components.
NPR corresponds to the normalized error
after projecting the fully sampled data onto the space spanned
either by the spatiotemporal maps or the PSF temporal functions.
For the STM model we define
\be
\mathrm{NPR}(L) \defequ 
\norm{\vrho - \hat{\vrho}(L)}_2/\norm{\vrho}_2,
\ee{e, NPR}
where
$\vrho$
denotes the reference image reconstructed from fully sampled data
and
$\hat{\vrho}(L)$ is calculated using \eref{e,est_rec}
with no regularization,
assuming an STM model with $L$ components,
and no undersampling operation in the system matrix \A.
Due to the high level of background noise in dataset (A.1),
we calculated the NPR
over a region of interest (ROI)
around the stomach
as illustrated
in Fig.~S2 of the supplementary material.

\subsection{Sketched SVD for a Projector Matrix Calculation: 
Evaluation of Computation Time and Representation Quality}

We tested the method proposed in Sec. \ref{sec:est_stm_sketch}
using datasets (A.1) and (B). 
Given that a random matrix is used, 
it is relevant to study 
whether different realizations can
induce a high variability in the final STM computation.
We varied the sketch dimension $s$ in the range $[2\rankC, 6\rankC]$,
and calculated the NPR for different realizations of the sketch matrix;
we reported the mean and standard deviation
using $50$ realizations in each case.
We also report the median time to compute 
an SVD with and without sketching.

\subsection{Reconstruction Experiments}

We tested the reconstruction framework in Sec. \ref{sec:rec}
using datasets (A.1), (A.2) and (B).
For datasets (A.1) and (B) we simulated accelerated acquisitions
by retrospectively undersampling the data
using the undersampling k-space masks described in Sec. \ref{sec:exp_mask}.
We reconstructed each accelerated dataset
using the proposed reconstruction framework where,
as a proof-of-principle, 
we explored Tikhonov regularization 
and
a P-LORAKS regularizer
\cite{haldar2016p},
where the latter
considered each STM spatial function as a virtual channel.
For simplicity, we refer to these two methods as
\stmTK
and
\stmLK
hereafter.
Our motivation for exploring a LORAKS-type regularizer
was based on our empirical observation 
that SILP relationships can be present
among the spatial components of the STM model,
and therefore structured low-rank models could be used \cite{haldar:20:lpi}. 
One theoretical explanation for the existence 
of these relationships could be based on
the shared spatial support of the spatial components
\cite{lobos2023extended}.
The two regularizers that we used were selected for simplicity,
as both involved only one regularization parameter. 
Alternatively,
one could use different regularizers for each spatial function
in the STM decomposition,
with individual regularization parameters.

When using Tikhonov regularization,
we minimized \eref{e,rec_inv}
using conjugate gradient.
When using P-LORAKS regularization,
we solved the inverse problem
using the majorize-minimize algorithm in \cite{kim2017loraks}.
We compared the reconstruction results
with five baseline methods.
The first
method used data sharing 
to substitute each missing k-space sample
with a sample available 
in a neighboring time frame \cite{jones1993k}.
We refer to this method as Data Sharing.
The second method used a locally low-rank (LLR) model 
that was implemented using nonlinear conjugate gradient \cite{lobos2025smooth}.
The third method corresponded to blind compressed sensing (BCS) \cite{lingala2013blind},
using the software provided by the authors.
The fourth method
used a low-rank plus sparsity model 
\cite{otazo:15:lrp}
that was implemented using POGM
\cite{lin:19:edp}.
We refer to this  method as \LpS.
The fifth method corresponded to an
analogous version of the reconstruction framework in Sec.~\ref{sec:rec},
where instead of using the STM model in \eref{e,est_rec}
and computing the STM spatial functions in \eref{e,rec_inv}, 
we used the PSF model and computed the PSF spatial functions, respectively. 
The PSF temporal functions were computed 
from the same ACS data used to compute spatiotemporal maps, 
and we used Tikhonov regularization when computing the PSF spatial functions.
We refer to this method as \psfTK hereafter.

We studied how the parameters related 
to the spatial resolution of spatiotemporal maps
affected reconstruction performance.
Using dataset (B)
we studied how reconstruction performance
varied when choosing the radius parameter $R \in \{2, 3, 4, 5\}$.
In addition, we reconstructed the retrospectively undersampled data
using spatiotemporal maps computed after interpolating
$12.5\%$, $25\%$, and $40\%$
of their k-space using the FFT-based PISCO approach.

For datasets (A.1) and (B) we report the 
normalized root mean square error (NRMSE)
in each case;
for dataset (A.1) the NRMSE was calculated 
over the same ROI used when calculating the NPR.
For dataset (B) t-score maps were calculated to assess functional activity.
For the multichannel experiments 
we calculated sensitivity maps using PISCO
\cite{lobos2023new}
and the ACS data of the first time frame.
Each method was implemented in-house using MATLAB R2023a
on a local server with an 
Intel Zeon Silver 4216 2.10 GHz CPU
and
263 GB RAM.
Open-source code
for computing STMs
is available at:
\url{https://github.com/ralobos/STM_MRI.git}.
This code extends the PISCO software \cite{lobos2023software},
by including
the sketched SVD
and the orthogonal iteration 
computational methods.

\begin{figure}[t] 
\centering 
\includegraphics[width=\linewidth]{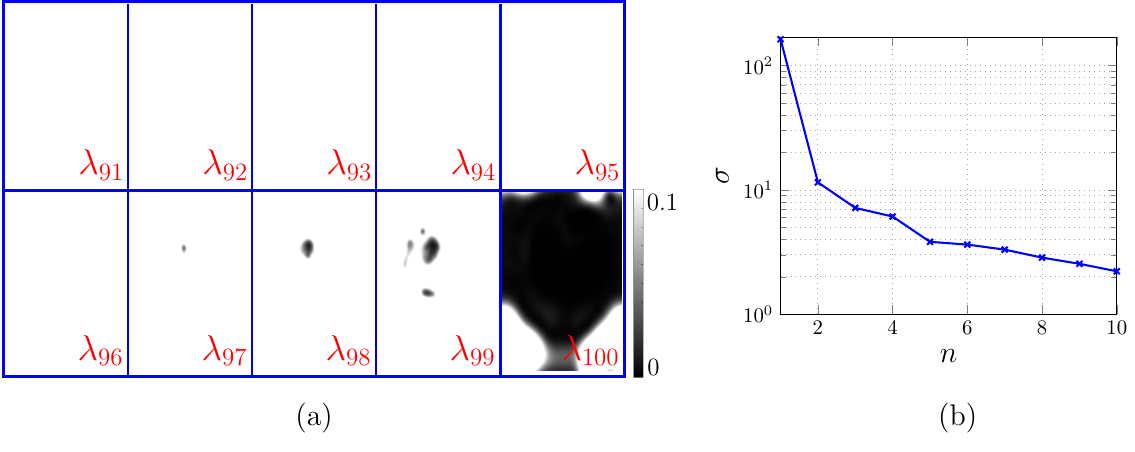}
\caption{(a) 
Eigenvalue maps representing the last $10$ eigenvalues 
(after 
normalizing and
sorting in decreasing order) 
of the matrices \Gx for each spatial location in the FOV
using dataset (A.1).
(b)
First $10$ singular values of the Casorati matrix
for dataset (A.1).
}
\label{fig:eig_svd}
\end{figure}
\begin{figure}[t] 
\centering 
\includegraphics[width=\linewidth]{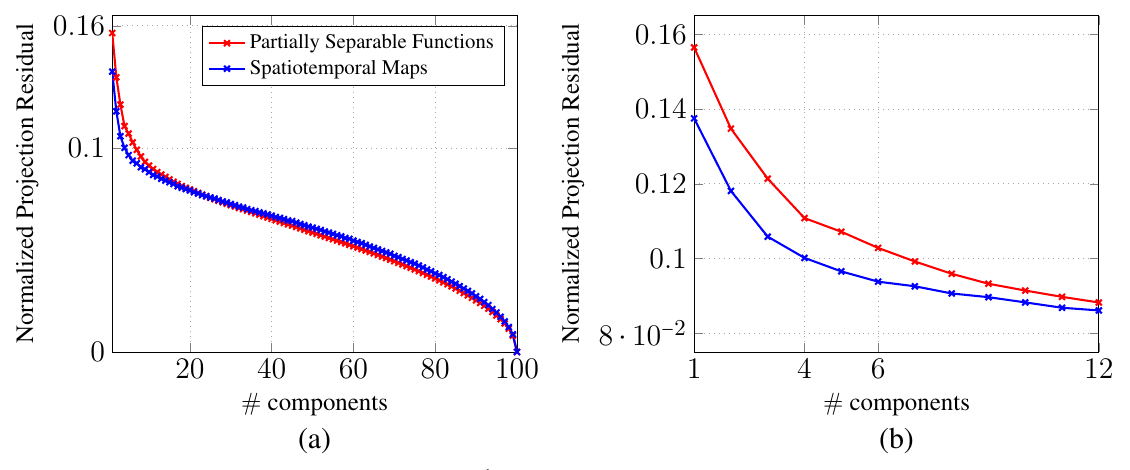}
\caption{
Normalized projection residual
versus the number of components
of the PSF and STM models.
(b) is a zoom in of (a).
}
\label{fig:rep_cap}
\end{figure}
\begin{figure}[t] 
\centering 
\includegraphics[width=\linewidth]{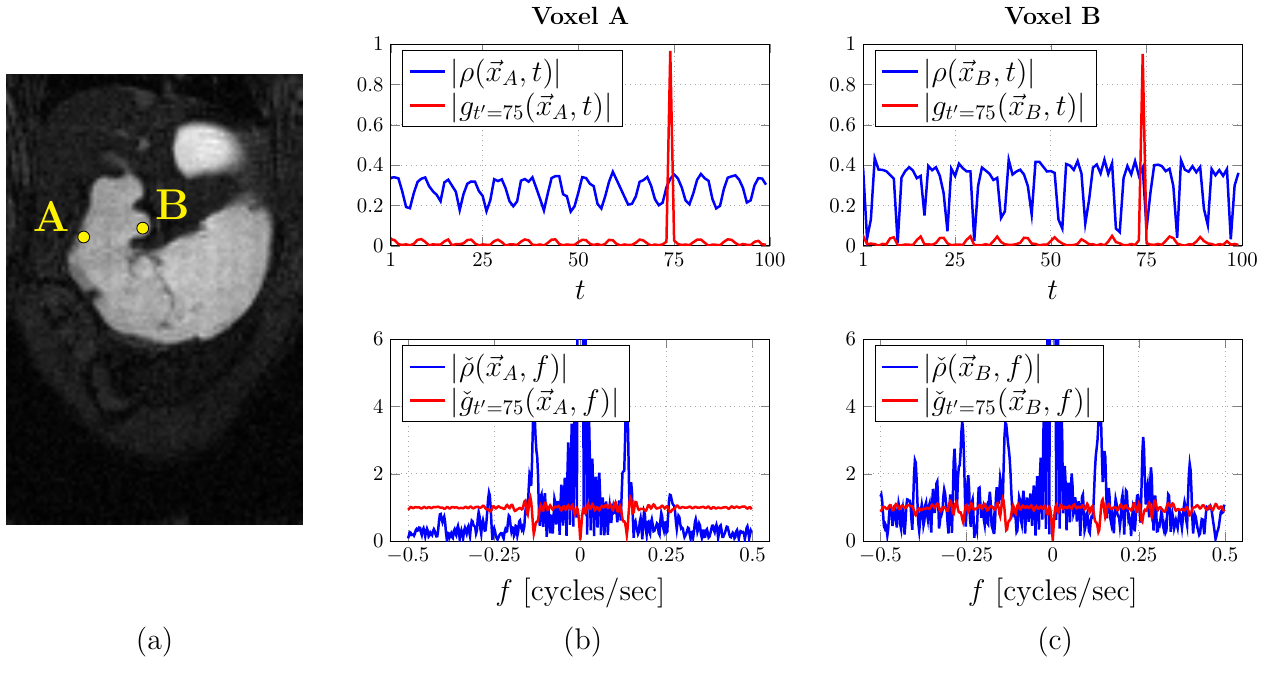}
\caption{
Visualization of the dynamic behavior of two voxels using dataset (A.1).
(a) Approximate locations of the two voxels 
overlaid with an image corresponding to the first time frame.
(b) Time evolutions of \rhoxt and \gvxt for voxel A (top),
and their respective frequency spectra (bottom). 
The frequency range is shown after normalization in cycles/sec.
(c) Analogous results for voxel B.
}
\label{fig:g_filt}
\end{figure}
\begin{figure}[t] 
\centering 
\includegraphics[width=0.45\textwidth]{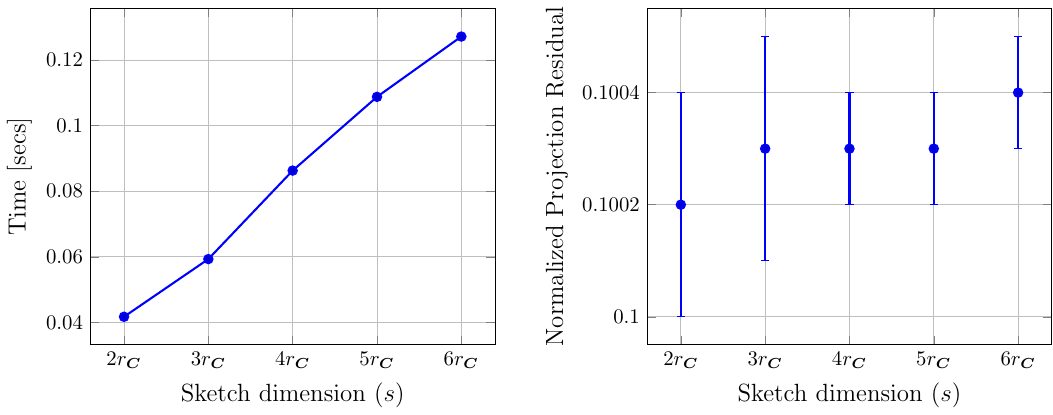}
\caption{
(a) Computation time of the sketched SVD
versus sketch dimension
for dataset (A.1). 
The median over $50$ realizations
is reported in each case.
(b) Mean and standard deviations of the NPR obtained in each case. 
}
\label{fig:time_NPR}
\end{figure}

\section{Results}
\label{sec:res}

Figure \ref{fig:eig_svd}.a shows
the last $10$ eigenvalues
(after sorting them in decreasing order)
of the \Gx matrices 
found when computing the STMs using  dataset~(A.1).
The number of eigenvalues near zero for each matrix \Gx
determines the number of components needed in the STM model,
as this is an estimation of the nullspace dimensionality.
The number of components varies across the FOV.
For example, areas that move slowly over time or that are motionless
and only present contrast variations, need only one component.
However, areas with more complex dynamics
like  
the antrum, which is in the distal part of the stomach,
need more components. 
The eigenvalue maps suggest
that four components could capture the dynamic behavior of these areas.
On the other hand,
the PSF model needs more components to represent 
the dynamic behaviors across the FOV.

Figure \ref{fig:eig_svd}.b shows the singular values
(sorted in decreasing order)
of the Casorati matrix used to find the PSF temporal functions.
The number of relatively large singular values
can be used to approximate the number of components needed in the PSF model
(\ie, \Lpsf);
\fref{fig:eig_svd}.b suggests that 4-6 
components should capture most of the dynamic behavior of the data.

Figure \ref{fig:rep_cap}
quantifies the representation error of both models
by measuring the NPR while varying the number of components
(\ie, $L$ in the STM model and $\Lpsf$ in the PSF model).
The NPR curve decreased faster
for the STM model than for the PSF model
when considering a small number of components.
The NPR for the STM model using $L=4$ components
(\ie, the number of components used in our experiments), 
was achieved by the PSF model when using at least $\Lpsf=6$ components.
Figure S1 shows qualitative results 
for the approximations given by the STM and PSF models
when $L = \Lpsf = 4$ for three representative time frames.
Each approximation was obtained after calculating 
the spatial components of each model from the fully sampled data.
Error images are also shown.
Overall, 
the STM approximation obtained smaller differences than the PSF approximation,
which is consistent with the NPR results in \fref{fig:rep_cap}.
The better representation capabilities of the  STM model 
can be attributed to the construction of different temporal subspaces 
for each spatial location.
The PSF model
construct a single temporal subspace
that is the same for all spatial locations.
Given that only 4 components were considered,
the single PSF temporal subspace 
was not able to properly represent the temporal behavior at every voxel.

Section \ref{sec:th_scond} provided sufficient conditions 
for the existence of 
SILP relationships in the \kt-space.
Using dataset (A.1),
we explored how these theoretical results 
can manifest in practice.
Specifically, we studied the SILP relationships
given by the multiframe FIR filters $\{\tgvkt\}_{t' =1}^T$
defined in \eref{e,g_filt},
which we calculated from the \Gx matrices.
Figure \ref{fig:g_filt} 
shows the temporal behavior of \rhoxt and \gvxt
for one specific $t'\in\{1,\ldots,T\}$,
considering two different voxels 
located in two areas with complicated dynamic behaviors.
It also shows \crhoxf and \cgvxf, 
from which we see that for both voxels 
\crhoxf has most of its energy in a finite set of bands
that depends on the spatial location of the voxel.
Furthermore, \cgvxf has its support defined such that
$\cgvxf\rhoxf \approx 0, \, \forall f \in \jcal F$,
which is related to the sufficient conditions provided in Sec. \ref{sec:th_scond}.

Table \ref{tb:time_NPR}
compares the compute time of
the regular SVD 
versus 
the sketched SVD
when computing spatiotemporal maps
using the 3D dataset (B).
The NPR obtained
in the final STM computation
is also shown in each case.
Both SVD approaches were applied 
on the matrix $\C^H\C$ whose dimensions were 
$35980 \times 35980$,
which caused regular SVD to be computationally expensive.
The sketched SVD was $\sim 640$-fold faster than the regular SVD,
reducing the computation from hours to seconds.
The spatiotemporal maps obtained for both cases
were quite similar
in terms of their representation capabilities,
which is reflected on the small differences in the NPR.
Both cases experience randomness from 
the random initialization used by the orthogonal iteration
when calculating the nullspace bases for the matrices \Gx.
This explains why the NPR result for regular SVD 
is shown with a standard deviation.

\begin{table}[h] 
\centering 
\caption{Compute time and 
representation quality
when using
sketched SVD and regular SVD 
for the computation of spatiotemporal maps
using the 3D dataset (B).
} 
\vspace{0.5cm} 
\begin{tabular}{|c|c|c|} 
\hline
\rowcolor{gray!30}
 & \textbf{Time [min]} & \textbf{NPR} \\ 
\hline
Regular SVD & 115.14 & 0.0565 $\pm$ 0.0001  \\ 
\hline
\rowcolor{gray!30}
Sketched SVD & 0.18 & 0.0575 $\pm$ 0.0003 \\ 
\hline
\end{tabular} 
\label{tb:time_NPR}
\end{table}

Given our heuristic approach to selecting the sketch dimension $s$,
we also explored the effects of varying its value.
Using dataset (A.1), \fref{fig:time_NPR} shows 
the computation times of using sketched SVD 
for different values of the sketched dimension,
and it also shows the NPR obtained in each case 
after computing the corresponding STMs.
Increasing the sketch dimension increased the computation time
as expected; 
however, 
STM representation quality
was only slightly affected
as the NPR displayed negligible fluctuations. 
In comparison to using regular SVD,
which took $5.77$ secs and obtained an NPR equal to $0.1005$,
a considerable acceleration was obtained
for each case considered in \fref{fig:time_NPR}.

\begin{figure}[t] 
\centering 
\includegraphics[width=0.4\textwidth]{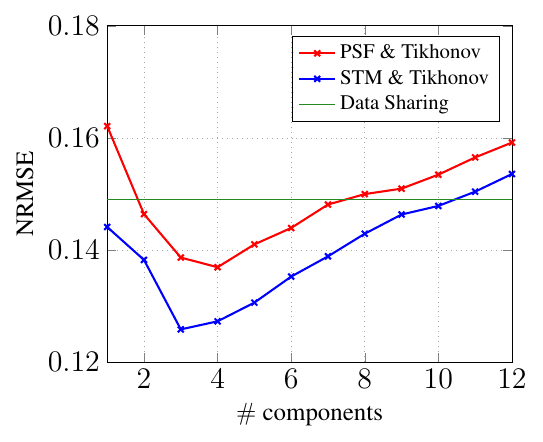}
\caption{
NRMSE 
versus the number of components
of the \psfTK and \stmTK methods.
The NRMSE obtained by Data Sharing is shown as a reference.
}
\label{fig:NRMSE_comps}
\end{figure}

\begin{figure*}[t] 
\centering 
\includegraphics[width=1\textwidth]{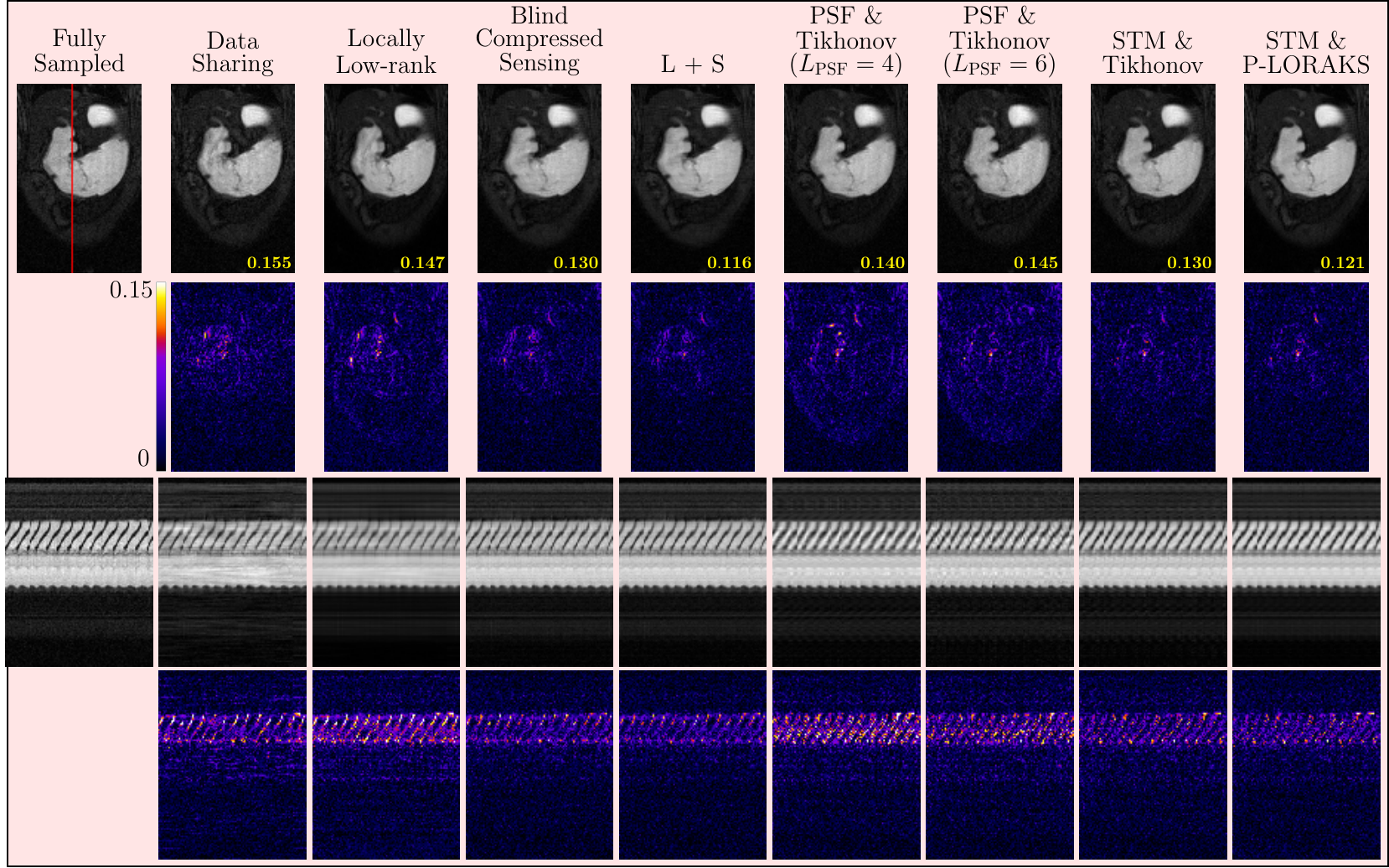}
\caption{
Retrospective reconstruction results using dataset (A.1).
The first time frame of the fully sampled data is shown on the left
next to a time profile of one line in the PE direction (indicated in red).
Then, reconstruction results are shown for each method, 
where the NRMSE for the displayed time frame
is indicated in yellow at the bottom right corner.
Error magnitude images are provided for each method
below the reconstructed images.
}
\label{fig:gmri_rec}
\end{figure*}

Figures~\ref{fig:NRMSE_comps} and~\ref{fig:gmri_rec} show
the reconstruction results using dataset (A.1).
Figure~\ref{fig:NRMSE_comps} shows the NRMSE obtained by 
\stmTK and \psfTK
when varying the number of components.
(For reference, the NRMSE obtained by Data Sharing is also shown.)
The NRMSE for both methods decreased initially as the number of components increased, 
as expected because the representation capabilities of both models
improve when increasing the number of components. 
The NRMSE achieved a minimum for both methods and then increased.
This is also expected as the number of unknowns increases as more components are included.
For all cases \stmTK obtained a lower NRMSE than \psfTK, 
which can be attributed to the better representation capabilities 
of the STM model compared to the PSF model (\cf, \fref{fig:rep_cap}).

Figure \ref{fig:gmri_rec} shows the reconstruction results 
for the methods based on the STM model when setting the number of components to $L=4$,
\psfTK when the number of components is set to $\Lpsf = 4$ and $\Lpsf = 6$,
and the other considered reconstruction methods.
The idea behind the considered values for \Lpsf
was to compare \stmTK and \psfTK when the number of components was the same,
and also when the latter had more components than the former. 
The first time frame is shown
for each reconstruction method, 
as well as 
one line in the PE direction for all the time frames.
This line covers several areas with a complex dynamic behavior,
and its evolution over time shows
how several voxels exhibit different temporal characteristics.
For the first time frame,
the regularized methods
all obtained similar NRMSE values
that were lower than the one obtained by Data Sharing
(indicated in \fref{fig:gmri_rec}).
However,
when considering the whole time series, 
\LpS and \stmLK
obtained better NRMSE than \stmTK and \psfTK
as reported in Table \ref{tb:time_NRMSE}.
Although 
\LpS obtained a slightly lower time-series NRMSE than \stmLK,
note that this value was calculated
with respect to a fully sampled dataset
that exhibited a high level of noise.
A method with denoising characteristics
can obtain a higher time-series NRMSE
in comparison to methods that preserve noise.
In our case, 
\stmLK had
better qualitative denoising characteristics
as seen in supplementary video~S1.
However, 
higher quantitative errors can be expected 
for voxels where noise was decreased. 
Figure S3 shows the reconstructed time course for voxel B in \fref{fig:g_filt}.a.
Even though \stmLK obtained a higher NRMSE than BCS or  \LpS for that specific voxel 
(\cf, Table S1),
it was able to better follow the intensity time profile of the fully sampled data.

\begin{figure*}[t] 
\centering 
\includegraphics[width=1\textwidth]{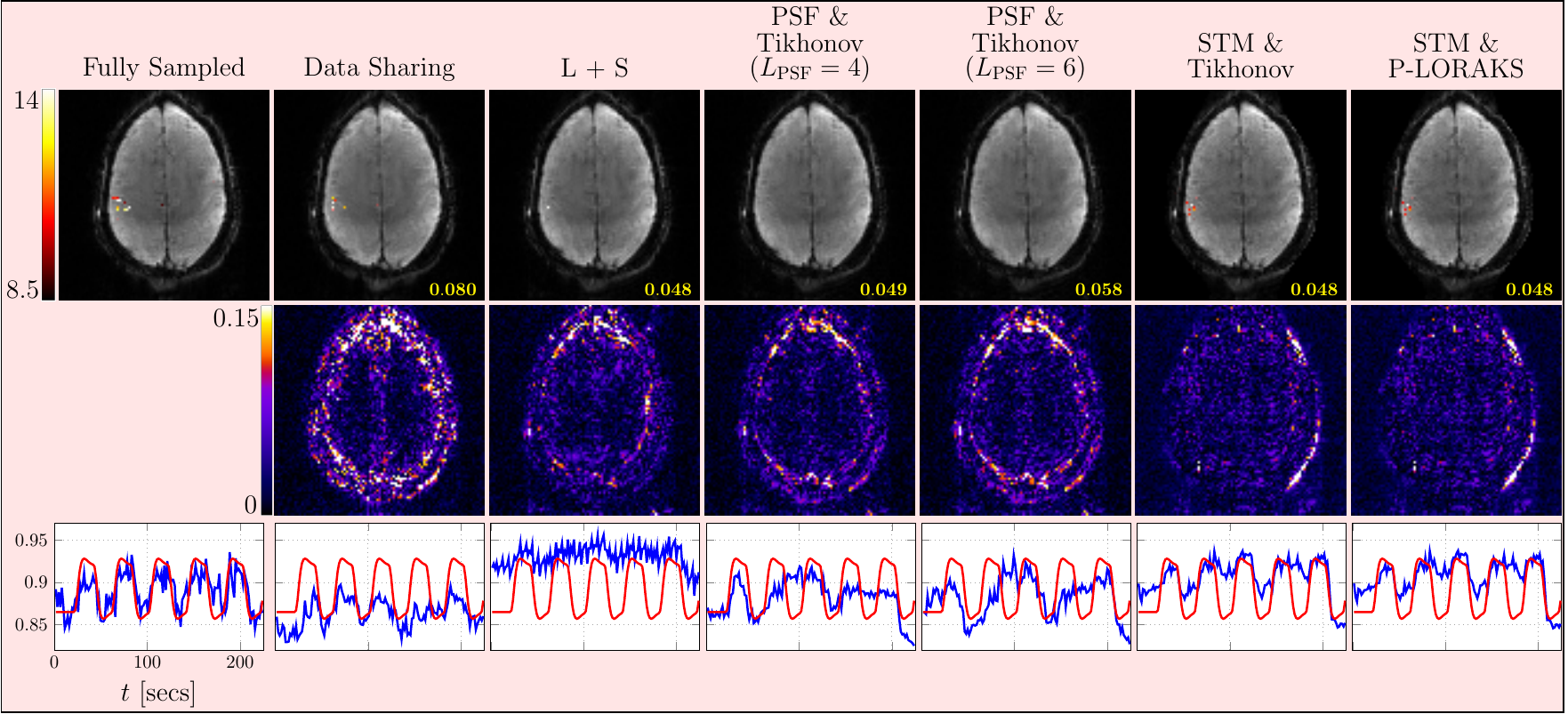}
\caption{
Retrospective reconstruction results using dataset (B).
Row 1:
The first time frame 
of a representative slice
of the fully sampled data is shown on the top left
overlaid with a t-score map calculated using the task information,
followed by reconstruction results for each method;
the NRMSE for the displayed time frame
is indicated in yellow at the bottom right corner.
The t-score maps obtained for each reconstruction method
are overlaid on the reconstructed images.
Row 2:
Error magnitude images are provided for each method.
Row 3:
Signal intensity plots;
the first plot corresponds to the signal evolution for one specific voxel 
with a high t-score using the fully sampled data (in blue), 
and a task curve (in red) showing the tap/rest blocks.
Analogous plots are also shown for all the reconstructed datasets
for the same voxel location.
}
\label{fig:fmri_rec}
\end{figure*}

Figure \ref{fig:fmri_rec} shows
the reconstruction results using the 3D dataset (B)
for the  same methods shown in \fref{fig:gmri_rec},
except for BCS and LLR\footnote{ 
The implementation of BCS for the 3D case
was not available, 
and extending the provided 2D implementation
required a nontrivial selection of parameters
that was beyond the scope of this paper;
LLR for the 3D case required a prohibitive computation time,
mainly due to the high data dimensions 
and the number of shifts needed such that overlapping patches
could be processed to avoid blocky artifacts \cite{lobos2025smooth}.
}.
In this case \stmLK and \stmTK obtained a similar time-series NRMSE,
which was better than those obtained by the other methods
(\cf, Table \ref{tb:time_NRMSE}),
and all methods produced quite similar qualitative results.
In addition to image quality, 
it was also important to capture functional activity.
Figure \ref{fig:fmri_rec}
shows
the t-score maps obtained for each reconstruction method
overlaid on the corresponding reconstructed images.
Even though the reconstructed images for each method exhibited good quality,
only \stmLK and \stmTK were able to obtain
t-score maps reflecting functional activity
similarly to the fully sampled data.
Figure \ref{fig:fmri_rec} also shows a plot with
the intensity over time
for one voxel with a high t-score
considering the fully sampled data 
and the reconstructed data for each method,
in addition to a ``task'' curve that shows the tap/rest blocks.
\stmLK and \stmTK
followed the task behavior 
better than \LpS
and Data Sharing.
The voxel time-series for \psfTKIV and \psfTKVI
were quite different than the task behavior.
Table S1 reports the NRMSE for the time course obtained by each reconstruction method. 
For this particular voxel,
methods using STMs obtained better values 
than \LpS,
and similar values to methods based on PSF.

We observed (not shown)
that the t-score maps for \LpS were improved
when using low acceleration factors; 
however, for high accelerations
a high degree of global low-rank regularization was needed 
to obtain good reconstruction quality,
causing the final reconstruction to have 
reduced temporal variability.
On the other hand, 
the STM model encodes dynamics at a voxel level,
which better captures phenomena that occur in a voxel-wise fashion,
as happens in fMRI.

Figure S4 shows reconstruction results for dataset (B) 
using \stmLK  while varying the radius parameter $R$ 
when computing STMs. 
Given that $R$ and the degree of interpolation
affect the spatial resolution of STMs,
we used no interpolation to study
the isolated effect of the former in the reconstruction.
Even though the time-series NRMSE was similar in each case
as shown in Table S2,
high values of $R$ better captured functional activity
than low values, 
as shown by the corresponding t-score maps
and the time plots of one specific voxel with a high t-score.
This is expected as higher values of $R$ allows the computation of STMs
with a higher spatial resolution that better capture 
abrupt changes in dynamic behavior
across spatial locations.
Figure S5 shows representative STMs corresponding to various values of $R$
for a specific time frame and the slice shown in \fref{fig:fmri_rec}.
As illustrated, increasing the value of $R$
resulted in less smooth STMs.
However, larger values of $R$ 
increase computational complexity and memory usage
and require larger autocalibration regions (\cf, Sec. III.A).

Table S3 shows the time-series NRMSE 
when reconstructing dataset (B) using \stmLK
and different degrees of interpolation 
when computing STMs. 
Similar quantitative results were obtained in each case.
Even though similar image quality 
and time-series NRMSE were obtained for each case
(\cf Table S3),
variations in functional activity were observed as shown in Fig.~S6.
The reconstructed time courses had higher variations
with respect to the task curve for higher interpolation degrees. 
The dynamic behavior in dataset (B) is expected to be localized
in brain regions related to the task,
and non-smooth variations in the dynamic behavior 
might be expected when transitioning
to voxels outside these regions.
This can require STMs
that are not necessarily smooth in those regions,
which decreases the degree of interpolation 
to accurately capture the dynamic behavior
of voxels activated during the task.

Supplementary video S2 shows the reconstruction results 
for the prospectively undersampled dataset (A.2).
Qualitatively the reconstructions 
were similar to the ones obtained 
when reconstructing restrospectively undersampled data using dataset (A.1).
BCS, \LpS, \stmTK, and \stmLK exhibited less aliasing artifacts
than the other methods. 
Methods based on STM exhibited less noise than BCS and \LpS;
although, sharper edges were observed for BCS and \LpS.

\begin{table}[h] 
\centering 
\caption{Time-series NRMSE for both datasets.
} 
\vspace{0.5cm} 
\begin{tabular}{|c|c|c|} 
\hline
\rowcolor{gray!30}
 & \textbf{2D dataset (A.1)}& \textbf{3D dataset (B)} \\ 
\hline
Zero filled & 0.200 & 0.346 \\ 
\hline
\rowcolor{gray!30}
Data Sharing & 0.149 & 0.206 \\
\hline
Locally Low-rank & 0.133 & n/a \\
\hline
\rowcolor{gray!30}
Blind Compressed Sensing & 0.119 & n/a  \\ 
\hline
\LpS & 0.109 & 0.116  \\ 
\hline
\rowcolor{gray!30}
\psfTK ($\Lpsf = 4$) & 0.137 & 0.124  \\ 
\hline
\psfTK ($\Lpsf = 6$) & 0.144 & 0.127  \\ 
\hline
\rowcolor{gray!30}
STM $\&$ Tikhonov  & 0.127 & 0.097 \\ 
\hline
STM $\&$ P-LORAKS & 0.113 & 0.097 \\ 
\hline
\end{tabular} 
\label{tb:time_NRMSE}
\end{table}

\section{Discussion and Conclusions}
\label{sec:dis_con}

The STM model proposed in this work for dynamic MRI reconstruction
has been presented as an extension of the widely used PSF model.
Unlike the PSF model, 
the STM model represents the dynamic MRI signal
using temporal functions
(\ie, spatiotemporal maps)
that depend on the spatial location.
We have shown that this spatial dependency
allows dynamic MRI reconstruction methods based on the STM model
to obtain better reconstruction than methods based on the PSF model
(\cf, Figs. \ref{fig:NRMSE_comps}, \ref{fig:gmri_rec}, \ref{fig:fmri_rec}
and Table \ref{tb:time_NRMSE}).

Another feature of the STM model is that 
the number of components can depend on the spatial location.
We have not exploited this flexibility
in the experiments shown in this work;
however, this spatial dependency could be used,
for example,
to focus the reconstruction on regions of interest. 
In particular, we hypothesize that spatiotemporal maps
could have a synergistic effect when combined with
Region-Optimized Virtual (ROVir) coils 
\cite{kim:21:rov}. 
 
The number of components in the STM model 
vary depending on the dynamics present in the data. 
In our experiments we observed that some quasi-periodic behaviors
related to small deformations or contrast changes
were well-approximated using a few components. 
However, we have observed (not shown) that 
more complex dynamics 
(\eg, respiratory motion) 
require several components. 
Our future work includes extending the STM framework
to more challenging contexts, 
and theoretically studying
how the number of needed components 
varies depending on the data dynamics.

We also provided a theoretical framework 
that connects the properties of spatiotemporal maps
with autoregressive properties of the \kt-space.
Specifically, we have shown that establishing 
shift-invariant linear predictability relationships
in the k-space of different time frames,
leads to a subspace-based computation framework for spatiotemporal maps.
Notably, our theory
allows a direct connection with 
subspace-based estimation of sensitivity maps in multichannel MRI.
Using this analogue, 
we provided a spatiotemporal map construction procedure
that relies on recent advanced signal processing computational methods, 
originally proposed for sensitivity map estimation.
Moreover, we have extended these computational methods
using randomized linear algebra techniques
to make spatiotemporal map computation
efficient for long time series.
As an example, 
using 3D fMRI data,
the computational method proposed in this work based on sketched SVD
was $\sim 640 \times $ faster
than using the analogous computational method
based on the previous SVD approach
originally proposed for sensitivity map estimation.
Even though our proposed computational methods 
were used in the context of STM computation,
they could also be used for 
subspace-based sensitivity map estimation
\cite{lobos2023new, uecker2014},
where compute time could be greatly reduced
for receiver arrays having many coils.

We have shown that spatiotemporal maps can be incorporated in the signal model, 
supporting numerous reconstruction frameworks
for accelerated dynamic MRI.
As proof-of-principle illustrations,
we investigated a model-based reconstruction framework
that includes spatiotemporal maps as part of a data-consistency term,
where they encode the information about the dynamics of the data.
The resulting inverse problem considers 
the spatial functions of the STM model as the optimization variables.
This formulation reduces the reconstruction
of a long dynamic series of images
to the estimation of a few non-dynamic component images.
Therefore, regularizers developed for non-dynamic MRI can be used. 
We evaluated the proposed reconstruction framework
using a simple Tikhonov regularizer 
and 
a LORAKS-type regularizer.
Using highly accelerated data,
both regularizers obtained similar
NRMSE metrics
as the \LpS method based on low-rank models.
However, the STM reconstruction framework showed
additional denoising characteristics, 
and better captured
small dynamic changes
that are particularly relevant in fMRI. 
As an example, 
t-scores maps produced
from fMRI images reconstructed using the STM-based approach
clearly revealed functional activity in expected spatial areas,
similarly to the gold standard; this was not the case 
when using a reconstruction based on a low-rank + sparsity model
or other methods based on the PSF model.
Interestingly, 
a simple data-sharing reconstruction exhibited a periodic pattern similar 
to that of the gold standard. 
A weighted version of this method could improve reconstruction performance
and serve as initialization for other methods.
We emphasize that
the regularizers used in our experiments are not necessarily optimal,
and they were just used as a proof-of-principle. 
Another option
would be to use 
regularizers based on machine/deep learning models
that were originally proposed in non-dynamic or dynamic settings.
Moreover, given that spatiotemporal maps
can be incorporated as part of the system forward operator,
they could be used in other reconstruction frameworks
that leverage on scan-specific machine learning methods,
\eg,
\cite{yu2025bilevel}.
Exploring these ideas is part of our future work.

The computation time for reconstruction frameworks based on the STM model 
can vary depending on how these are employed. 
In this work, we have presented a proof-of-principle, 
model-based reconstruction framework,
where the total computation time includes both STM calculation
and the time spent solving the optimization problem--which itself 
depends on the choice of regularizer.  
STM computation can be performed efficiently 
using PISCO and our proposed sketched SVD technique (\cf, Sec. \ref{sec:est_stm}).
For instance, STM computation required only $\sim  5$ secs
for dataset (A.1).
However, solving the optimization problem in \eref{e,rec_inv}
using iterative methods 
necessitates performing a number of FFTs equal to the number of time frames.
For long time series, this process can be time-intensive.
This can be considered as a disadvantage compared to
reconstruction frameworks based on the PSF model. 
As shown in certain applications, 
the PSF model's structure can be exploited
to reduce the number of required FFTs--from the total number of time frames down to the number of spatial components \cite{mani:15:fia}.
Additionally, the STM model requires more memory to store its parameters: 
a different set of temporal functions needs to be stored for each voxel, 
in contrast to the PSF model
that requires only a single set of temporal functions for all voxels.
In future work, 
we plan to investigate 
more efficient implementations for reconstruction frameworks based on the STM model.

Our future work also includes using the STM model
in other applications.
For example, 
given the existence of linear predictability in the \kt-space
of images with different contrast 
\cite{haldar:20:lpi, bilgic2018improving},
the STM model could be used
to reconstruct
accelerated quantitative MRI data (\cf, Sec. \ref{sec:SILP_stm}).


\bibliographystyle{IEEEtran}
\bibliography{./bib-jf,./bib-rl}

\providecommand{\noopsort}[1]{}
\begin{thebibliography}{10}
\providecommand{\url}[1]{#1}
\csname url@samestyle\endcsname
\providecommand{\newblock}{\relax}
\providecommand{\bibinfo}[2]{#2}
\providecommand{\BIBentrySTDinterwordspacing}{\spaceskip=0pt\relax}
\providecommand{\BIBentryALTinterwordstretchfactor}{4}
\providecommand{\BIBentryALTinterwordspacing}{\spaceskip=\fontdimen2\font plus
\BIBentryALTinterwordstretchfactor\fontdimen3\font minus
  \fontdimen4\font\relax}
\providecommand{\BIBforeignlanguage}[2]{{%
\expandafter\ifx\csname l@#1\endcsname\relax
\typeout{** WARNING: IEEEtran.bst: No hyphenation pattern has been}%
\typeout{** loaded for the language `#1'. Using the pattern for}%
\typeout{** the default language instead.}%
\else
\language=\csname l@#1\endcsname
\fi
#2}}
\providecommand{\BIBdecl}{\relax}
\BIBdecl

\bibitem{jones1993k}
R.~Jones, O.~Haraldseth, T.~M{\"u}ller, P.~Rinck, and A.~{\O}ksendal, ``K-space
  substitution: a novel dynamic imaging technique,'' \emph{Mag{.} Res{.}
  Med{.}}, vol.~29, no.~6, pp. 830--834, Jun. 1993.

\bibitem{liang1997dynamic}
Z.-P. Liang, H.~Jiang, C.~P. Hess, and P.~C. Lauterbur, ``Dynamic imaging by
  model estimation,'' \emph{Int. J. Imag. Syst. Tech.}, vol.~8, no.~6, pp.
  551--557, 1997.

\bibitem{madore1999unaliasing}
B.~Madore, G.~H. Glover, and N.~J. Pelc, ``Unaliasing by {F}ourier-encoding the
  overlaps using the temporal dimension ({UNFOLD}), applied to cardiac imaging
  and f{MRI},'' \emph{Mag{.} Res{.} Med{.}}, vol.~42, no.~5, pp. 813--828,
  1999.

\bibitem{liang:07:siw}
Z.-P. Liang, ``Spatiotemporal imaging with partially separable functions,'' in
  \emph{{Proc. IEEE Intl. Symp. Biomed. Imag.}}, 2007, pp. {988--91}.

\bibitem{tsao2003k}
J.~Tsao, P.~Boesiger, and K.~P. Pruessmann, ``k-t {BLAST} and k-t {SENSE}:
  dynamic {MRI} with high frame rate exploiting spatiotemporal correlations,''
  \emph{Mag{.} Res{.} Med{.}}, vol.~50, no.~5, pp. 1031--1042, 2003.

\bibitem{jung2009k}
H.~Jung, K.~Sung, K.~S. Nayak, E.~Y. Kim, and J.~C. Ye, ``k-t {FOCUSS}: a
  general compressed sensing framework for high resolution dynamic mri,''
  \emph{Mag{.} Res{.} Med{.}}, vol.~61, no.~1, pp. 103--116, 2009.

\bibitem{haldar2010spatiotemporal}
J.~P. Haldar and Z.-P. Liang, ``Spatiotemporal imaging with partially separable
  functions: A matrix recovery approach,'' in \emph{Proc. IEEE Int. Symp.
  Biomed. Imaging}.\hskip 1em plus 0.5em minus 0.4em\relax IEEE, 2010, pp.
  716--719.

\bibitem{lingala2011accelerated}
S.~G. Lingala, Y.~Hu, E.~DiBella, and M.~Jacob, ``Accelerated dynamic mri
  exploiting sparsity and low-rank structure: kt slr,'' \emph{IEEE Trans. Med.
  Imag.}, vol.~30, no.~5, pp. 1042--1054, 2011.

\bibitem{zhao:12:irf}
B.~Zhao, J.~P. Haldar, A.~G. Christodoulou, and Z.-P. Liang, ``Image
  reconstruction from highly undersampled (k, t)-space data with joint partial
  separability and sparsity constraints,'' \emph{{IEEE Trans. Med. Imag.}},
  vol.~31, no.~9, pp. {1809--20}, Sep. 2012.

\bibitem{trzasko:11:lvg}
\BIBentryALTinterwordspacing
J.~Trzasko and A.~Manduca, ``Local versus global low-rank promotion in dynamic
  {MRI} series reconstruction,'' in \emph{{Proc. Intl. Soc. Mag. Res. Med.}},
  2011, p. 4371. [Online]. Available:
  \url{http://archive.ismrm.org/2011/4371.html}
\BIBentrySTDinterwordspacing

\bibitem{lingala2013blind}
S.~G. Lingala and M.~Jacob, ``Blind compressive sensing dynamic {MRI},''
  \emph{IEEE Trans. Med. Imag.}, vol.~32, no.~6, pp. 1132--1145, Mar. 2013.

\bibitem{otazo:15:lrp}
R.~Otazo, E.~{Candes}, and D.~K. Sodickson, ``Low-rank plus sparse matrix
  decomposition for accelerated dynamic {MRI} with separation of background and
  dynamic components,'' \emph{{Mag. Res. Med.}}, vol.~73, no.~3, pp.
  {1125--36}, Mar. 2015.

\bibitem{feng2016xd}
L.~Feng, L.~Axel, H.~Chandarana, K.~T. Block, D.~K. Sodickson, and R.~Otazo,
  ``{XD-GRASP}: golden-angle radial {MRI} with reconstruction of extra
  motion-state dimensions using compressed sensing,'' \emph{Mag{.} Res{.}
  Med{.}}, vol.~75, no.~2, pp. 775--788, 2016.

\bibitem{poddar2015dynamic}
S.~Poddar and M.~Jacob, ``Dynamic {MRI} using smoothness regularization on
  manifolds ({SToRM}),'' \emph{IEEE Trans. Med. Imag.}, vol.~35, no.~4, pp.
  1106--1115, 2015.

\bibitem{ravishankar:17:lra}
S.~Ravishankar, B.~E. Moore, R.~R. Nadakuditi, and J.~A. Fessler, ``Low-rank
  and adaptive sparse signal {(LASSI)} models for highly accelerated dynamic
  imaging,'' \emph{{IEEE Trans. Med. Imag.}}, vol.~36, no.~5, pp. {1116--28},
  May 2017.

\bibitem{lin:19:edp}
C.~Y. Lin and J.~A. Fessler, ``Efficient dynamic parallel {MRI} reconstruction
  for the low-rank plus sparse model,'' \emph{{IEEE Trans. Computational
  Imaging}}, vol.~5, no.~1, pp. {17--26}, Mar. 2019.

\bibitem{biswas2019dynamic}
S.~Biswas, H.~K. Aggarwal, and M.~Jacob, ``Dynamic {MRI} using model-based deep
  learning and {SToRM} priors: {MoDL-SToRM},'' \emph{Mag{.} Res{.} Med{.}},
  vol.~82, no.~1, pp. 485--494, 2019.

\bibitem{qin2018convolutional}
C.~Qin, J.~Schlemper, J.~Caballero, A.~N. Price, J.~V. Hajnal, and D.~Rueckert,
  ``Convolutional recurrent neural networks for dynamic {MR} image
  reconstruction,'' \emph{IEEE Trans. Med. Imag.}, vol.~38, no.~1, pp.
  280--290, 2018.

\bibitem{ke:21:llr}
Z.~Ke, W.~Huang, Z.-X. Cui, J.~Cheng, S.~Jia, H.~Wang, X.~Liu, H.~Zheng,
  L.~Ying, Y.~Zhu, and D.~Liang, ``Learned low-rank priors in dynamic {MR}
  imaging,'' \emph{{IEEE Trans. Med. Imag.}}, vol.~40, no.~12, pp. {3698--710},
  2021.

\bibitem{cruz2023low}
G.~Cruz, A.~Hua, C.~Munoz, T.~F. Ismail, A.~Chiribiri, R.~M. Botnar, and
  C.~Prieto, ``Low-rank motion correction for accelerated free-breathing
  first-pass myocardial perfusion imaging,'' \emph{Mag{.} Res{.} Med{.}},
  vol.~90, no.~1, pp. 64--78, 2023.

\bibitem{lobos2025smooth}
R.~A. Lobos, J.~S. Cavazos, R.~R. Nadakuditi, and J.~A. Fessler, ``Smooth
  optimization using global and local low-rank regularizers,'' \emph{arXiv
  preprint arXiv:2505.06073}, May 2025.

\bibitem{feng2025spatiotemporal}
J.~Feng, R.~Feng, Q.~Wu, X.~Shen, L.~Chen, X.~Li, L.~Feng, J.~Chen, Z.~Zhang,
  C.~Liu \emph{et~al.}, ``Spatiotemporal implicit neural representation for
  unsupervised dynamic {MRI} reconstruction,'' \emph{IEEE Trans. Med. Imag.},
  Jan. 2025.

\bibitem{yu2025bilevel}
H.~Yu, J.~A. Fessler, and Y.~Jiang, ``Bilevel optimized implicit neural
  representation for scan-specific accelerated {MRI} reconstruction,''
  \emph{arXiv preprint arXiv:2502.21292}, Feb. 2025.

\bibitem{manjon2013diffusion}
J.~V. Manj{\'o}n, P.~Coup{\'e}, L.~Concha, A.~Buades, D.~L. Collins, and
  M.~Robles, ``Diffusion weighted image denoising using overcomplete local
  {PCA},'' \emph{PloS one}, vol.~8, no.~9, p. e73021, 2013.

\bibitem{saucedo2017improved}
A.~Saucedo, S.~Lefkimmiatis, N.~Rangwala, and K.~Sung, ``Improved computational
  efficiency of locally low rank {MRI} reconstruction using iterative random
  patch adjustments,'' \emph{IEEE Trans. Med. Imag.}, vol.~36, no.~6, pp.
  1209--1220, 2017.

\bibitem{tamir:17:tss}
J.~I. Tamir, M.~Uecker, W.~Chen, P.~Lai, M.~T. Alley, S.~S. Vasanawala, and
  M.~Lustig, ``T2 shuffling: sharp, multicontrast, volumetric fast spin-echo
  imaging,'' \emph{{Mag. Res. Med.}}, vol.~77, no.~1, pp. {180--95}, Jan. 2017.

\bibitem{cordero2019complex}
L.~Cordero-Grande, D.~Christiaens, J.~Hutter, A.~N. Price, and J.~V. Hajnal,
  ``Complex diffusion-weighted image estimation via matrix recovery under
  general noise models,'' \emph{NeuroImage}, vol. 200, pp. 391--404, Oct. 2019.

\bibitem{guo:20:hro}
S.~Guo, J.~A. Fessler, and D.~C. Noll, ``High-resolution oscillating
  steady-state {fMRI} using patch-tensor low-rank reconstruction,'' \emph{{IEEE
  Trans. Med. Imag.}}, vol.~39, no.~12, pp. {4357--68}, Dec. 2020.

\bibitem{vizioli:21:ltt}
L.~Vizioli, S.~Moeller, L.~Dowdle, M.~{Akcakaya}, F.~D. Martino, E.~Yacoub, and
  K.~{Ugurbil}, ``Lowering the thermal noise barrier in functional brain
  mapping with magnetic resonance imaging,'' \emph{{Nature Comm.}}, vol.~12, p.
  5181, 2021.

\bibitem{zhao2022joint}
Y.~Zhao, Z.~Yi, L.~Xiao, V.~Lau, Y.~Liu, Z.~Zhang, H.~Guo, A.~T. Leong, and
  E.~X. Wu, ``Joint denoising of diffusion-weighted images via structured
  low-rank patch matrix approximation,'' \emph{Mag{.} Res{.} Med{.}}, vol.~88,
  no.~6, pp. 2461--2474, Dec. 2022.

\bibitem{comby:23:dof}
P.-A. Comby, Z.~Amor, A.~Vignaud, and P.~Ciuciu, ``Denoising of {fMRI} volumes
  using local low rank methods,'' in \emph{{Proc. IEEE Intl. Symp. Biomed.
  Imag.}}, 2023, pp. {1--5}.

\bibitem{meyer:23:ect}
N.~K. Meyer, D.~Kang, D.~F. Black, N.~G. Campeau, K.~M. Welker, E.~M. Gray,
  M.-H. In, Y.~Shu, J.~Huston, M.~A. Bernstein, and J.~D. Trzasko, ``Enhanced
  clinical task-based {fMRI} metrics through locally low-rank denoising of
  complex-valued data,'' \emph{{The Neuroradiology J.}}, vol.~36, no.~3, pp.
  {273--88}, Jun. 2023.

\bibitem{chen:23:isl}
X.~Chen \emph{et~al.}, ``Improved structured low-rank reconstruction for {3D}
  multi-shot {EPI} with joint motion modelling,'' in \emph{{ISMRM Workshop on
  Data Sampling and Image Reconstruction}}, 2023.

\bibitem{haldar:20:lpi}
J.~P. Haldar and K.~Setsompop, ``Linear predictability in {MRI} reconstruction:
  {Leveraging} shift-invariant {Fourier} structure for faster and better
  imaging,'' \emph{{IEEE Sig. Proc. Mag.}}, vol.~37, no.~1, pp. {69--82}, Jan.
  2020.

\bibitem{mishali2009blind}
M.~Mishali and Y.~C. Eldar, ``Blind multiband signal reconstruction:
  {C}ompressed sensing for analog signals,'' \emph{IEEE Trans. Signal
  Process.}, vol.~57, no.~3, pp. 993--1009, Mar. 2009.

\bibitem{lobos2023new}
R.~A. Lobos, C.-C. Chan, and J.~P. Haldar, ``New theory and faster computations
  for subspace-based sensitivity map estimation in multichannel {MRI},''
  \emph{IEEE Trans. Med. Imag.}, vol.~43, no.~1, pp. {286--96}, Jan. 2024.

\bibitem{lobos2023extended}
------, ``Extended version of ``{N}ew theory and faster computations for
  subspace-based sensitivity map estimation in multichannel {MRI}'',''
  \emph{arXiv preprint arXiv:2302.13431}, Feb. 2023.

\bibitem{lobos2023software}
\BIBentryALTinterwordspacing
------, ``{PISCO} software version 1.0,'' University of Southern California,
  Los Angeles, CA, Tech. Rep. USC-SIPI-458, Mar. 2023. [Online]. Available:
  \url{https://sipi.usc.edu/reports/abstracts.php?rid=sipi-458}
\BIBentrySTDinterwordspacing

\bibitem{gilbert2012sketched}
A.~C. Gilbert, J.~Y. Park, and M.~B. Wakin, ``Sketched {SVD}: {R}ecovering
  spectral features from compressive measurements,'' \emph{arXiv preprint
  arXiv:1211.0361}, Nov. 2012.

\bibitem{golub2013}
G.~Golub and C.~van Loan, \emph{Matrix Computations}, 4th~ed.\hskip 1em plus
  0.5em minus 0.4em\relax The Johns Hopkins University Press, 2013.

\bibitem{lobos2025stm_ismrm}
R.~A. Lobos, X.~Wang, Z.~Liu, J.~A. Fessler, and D.~C. Noll, ``Spatiotemporal
  maps for dynamic {MRI} reconstruction: a proof-of-principle demonstration on
  single-coil animal gastrointestinal data,'' in \emph{Proc. Int. Soc. Magn.
  Reson. Med.}, 2025, p. 2622.

\bibitem{uecker2014}
M.~Uecker, P.~Lai, M.~J. Murphy, P.~Virtue, M.~Elad, J.~M. Pauly, S.~S.
  Vasanawala, and M.~Lustig, ``{ESPIRiT} -- an eigenvalue approach to
  autocalibrating parallel {MRI}: Where {SENSE} meets {GRAPPA},'' \emph{Mag{.}
  Res{.} Med{.}}, vol.~71, pp. 990--1001, Mar. 2014.

\bibitem{haldar:14:lrm}
J.~P. Haldar, ``Low-rank modeling of local k-space neighborhoods {(LORAKS)} for
  constrained {MRI},'' \emph{{IEEE Trans. Med. Imag.}}, vol.~33, no.~3, pp.
  {668--81}, Mar. 2014.

\bibitem{haldar2016p}
J.~P. Haldar and J.~Zhuo, ``{P-LORAKS}: low-rank modeling of local k-space
  neighborhoods with parallel imaging data,'' \emph{Mag{.} Res{.} Med{.}},
  vol.~75, no.~4, pp. 1499--1514, Apr. 2016.

\bibitem{lobos:22:ots}
R.~A. Lobos and J.~P. Haldar, ``On the shape of convolution kernels in {MRI}
  reconstruction: {Rectangles} versus ellipsoids,'' \emph{{Mag. Res. Med.}},
  vol.~87, no.~6, pp. {2989--96}, Jun. 2022.

\bibitem{vershynin2018high}
R.~Vershynin, \emph{High-dimensional probability: {A}n introduction with
  applications in data science}.\hskip 1em plus 0.5em minus 0.4em\relax
  Cambridge {U}niversity {P}ress, 2018.

\bibitem{strang:93:tft}
G.~Strang, ``The fundamental theorem of linear algebra,'' \emph{{Amer. Math.
  Monthly}}, vol. 100, no.~9, pp. {848--55}, 1993.

\bibitem{fessler2024linear}
J.~A. Fessler and R.~R. Nadakuditi, \emph{Linear {A}lgebra for {D}ata
  {S}cience, {M}achine {L}earning, and {S}ignal {P}rocessing}.\hskip 1em plus
  0.5em minus 0.4em\relax Cambridge {U}niversity {P}ress, 2024.

\bibitem{pruessmann:99:sse}
K.~P. Pruessmann, M.~Weiger, M.~B. Scheidegger, and P.~Boesiger, ``{SENSE:}
  sensitivity encoding for fast {MRI},'' \emph{{Mag. Res. Med.}}, vol.~42,
  no.~5, pp. {952--62}, Nov. 1999.

\bibitem{kim2017loraks}
T.~H. Kim, K.~Setsompop, and J.~P. Haldar, ``{LORAKS} makes better {SENSE}:
  phase-constrained partial fourier {SENSE} reconstruction without phase
  calibration,'' \emph{Mag{.} Res{.} Med{.}}, vol.~77, no.~3, pp. 1021--1035,
  Mar. 2017.

\bibitem{wang2023diffeomorphic}
X.~Wang, J.~Cao, K.~Han, M.~Choi, Y.~She, U.~M. Scheven, R.~Avci, P.~Du, L.~K.
  Cheng, M.~R. Di~Natale \emph{et~al.}, ``Diffeomorphic surface modeling for
  {MR}i-based characterization of gastric anatomy and motility,'' \emph{IEEE
  Trans. Biomed. Eng.}, vol.~70, no.~7, pp. 2046--2057, Jan. 2023.

\bibitem{buehrer:07:acf}
M.~Buehrer, K.~P. Pruessmann, P.~Boesiger, and S.~Kozerke, ``Array compression
  for {MRI} with large coil arrays,'' \emph{{Mag. Res. Med.}}, vol.~57, no.~6,
  pp. {1131--9}, Jun. 2007.

\bibitem{xiang2024model}
H.~Xiang, J.~A. Fessler, and D.~C. Noll, ``Model-based reconstruction for
  looping-star {MRI},'' \emph{Mag{.} Res{.} Med{.}}, vol.~91, no.~5, pp.
  2104--2113, May 2024.

\bibitem{kim:21:rov}
D.~Kim, S.~F. Cauley, K.~S. Nayak, R.~M. Leahy, and J.~P. Haldar,
  ``Region-optimized virtual {(ROVir)} coils: {Localization} and/or suppression
  of spatial regions using sensor-domain beamforming,'' \emph{{Mag. Res.
  Med.}}, vol.~86, no.~1, pp. {197--212}, Jul. 2021.

\bibitem{mani:15:fia}
M.~Mani, M.~Jacob, V.~Magnotta, and J.~Zhong, ``Fast iterative algorithm for
  the reconstruction of multi-shot {non-Cartesian} diffusion data,''
  \emph{{Mag. Res. Med.}}, vol.~74, no.~4, pp. {1086--94}, Oct. 2015.

\bibitem{bilgic2018improving}
B.~Bilgic, T.~H. Kim, C.~Liao, M.~K. Manhard, L.~L. Wald, J.~P. Haldar, and
  K.~Setsompop, ``Improving parallel imaging by jointly reconstructing
  multi-contrast data,'' \emph{Mag{.} Res{.} Med{.}}, vol.~80, no.~2, pp.
  619--632, Jan. 2018.

\end{thebibliography}

\end{document}